\documentclass[acmsmall]{acmart}

\usepackage[ruled]{algorithm2e} 
\usepackage{csquotes}
\usepackage{multirow}
\usepackage{rotating}  
\usepackage{hyperref}
\usepackage{amsmath} 
\usepackage{amssymb}  
\usepackage{graphicx}
\usepackage{subcaption}
\usepackage{color, colortbl}
\usepackage{booktabs}

\SetAlFnt{\small}
\SetAlCapFnt{\small}
\SetAlCapNameFnt{\small}
\SetAlCapHSkip{0pt}
\IncMargin{-\parindent}
\usepackage{ctable} 
\definecolor{Gray}{gray}{0.9}

\def\hlinewd#1{%
\noalign{\ifnum0=`}\fi\hrule \@height #1 %
\futurelet\reserved@a\@xhline} 

\AtBeginDocument{%
  \providecommand\BibTeX{{%
    \normalfont B\kern-0.5em{\scshape i\kern-0.25em b}\kern-0.8em\TeX}}}

\setcopyright{rightsretained}
\copyrightyear{2020}
\acmYear{2020}
\acmDOI{00.0000/0000000.0000000}

\acmJournal{JACM}
\acmVolume{0}
\acmNumber{0}
\acmArticle{000}
\acmMonth{0}

\begin{document}

\title{Human-Collective Collaborative Site Selection}

\author{Jason R. Cody}
\orcid{1234-5678-9012-3456}
\affiliation{%
	\institution{United States Military Academy}
	\streetaddress{606 Thayer Road}
	\city{West Point}
	\state{NY}
	\postcode{10996}
	\country{USA}}
\email{jason.cody@westpoint.edu}
\author{Karina A. Roundtree}
\affiliation{%
	\institution{Oregon State University}
	\city{Corvallis, Oregon}
	\country{USA}}
\author{Julie A. Adams}
\affiliation{%
	\institution{Oregon State University}
	\city{Corvallis, Oregon}
	\country{USA}}

\renewcommand{\shortauthors}{Cody, Roundtree, and Adams}

\begin{abstract}
 Robotic collectives are large groups ($\ge50$) of locally sensing and communicating robots that encompass characteristics of swarms and colonies, whose emergent behaviors accomplish complex tasks.  Future human-collective teams will extend the ability of operators to monitor, respond, and make decisions in disaster response, search and rescue, and environmental monitoring problems.  This manuscript evaluates two collective best-of-$n$ decision models for enabling collectives to identify and choose the highest valued target  from a finite set of $n$ targets.  Two challenges impede the future use of human-collective shared decisions:  1) environmental bias reduces collective decision accuracy when poorer targets are easier to evaluate than higher quality targets, and 2) little is understood about shared human-collective decision making interaction strategies.  The two evaluated collective best-of-$n$ models include an existing insect colony decision model and an extended bias-reducing model that attempts to reduce environmental bias in order to improve accuracy.  Collectives using these two strategies are compared independently and as members of human-collective teams.  Independently, the extended model is slower than the original model, but the extended algorithm is 57\% more accurate in decisions where the optimal option is more difficult to evaluate.  Human-collective teams using the bias-reducing model require less operator influence and achieve 25\% higher accuracy with difficult decisions, than the human-collective teams using the original model.  Further, a novel human-collective interaction strategy enables operators to adjust collective autonomy while making multiple simultaneous decisions.  
  
\end{abstract}

\ccsdesc[500]{Computing methodologies~Multi-agent systems}
\ccsdesc[300]{Human-centered computing~HCI theory, concepts and models}

\keywords{human-swarm interaction, collective decision making, swarm intelligence}

\maketitle

\section{Introduction}
\label{sec:HCI_Introduction}

The increasing capability and availability of robotic platforms continues to inspire the development of robotic collectives, which are large ($\ge 50$) groups of locally sensing and communicating robots that coordinate to solve problems \cite{Sahin2005}.  The term \textit{robotic collective} refers to robotic systems that exhibit behaviors analogous to the behaviors of biological spatial swarms (e.g., fish schools \cite{Couzin2005:1}) and colonies (e.g., ant and honeybee colonies \cite{PrattSumpter2002:InfoFlowOpinionPollCollectiveIntel}). Individual entities in spatial swarms perform similar tasks (e.g., coordinated movement \cite{Goodrich2011:1},  aggregation \cite{arvin2014,garnier2009}) and share local information  in a manner that causes the emergence of group behaviors, which are robust to individual failure, flexible in response to environmental changes, and scale in population size \cite{Brambilla2013:SwarmRoboticsReview}. Robotic colonies share the advantages of spatial swarms, but colonies are capable of advanced behaviors, including collective decision making \cite{PrattSumpter2002:InfoFlowOpinionPollCollectiveIntel}.  A colony's individual entities perform different tasks simultaneously and pool information, typically inside a decision making hub \cite{Valentini_Survey_2017}, similar to a honeybee hive \cite{Seeley1995:Wisdom}.  Information pooling enables the colony to make group decisions about current and future actions, as seen in biological colony foraging \cite{Seeley1986} and site selection \cite{PrattSumpter2002:InfoFlowOpinionPollCollectiveIntel}. Colony decision making hubs facilitate two different types of collective decisions: task allocation and consensus achievement \cite{Brambilla2013:SwarmRoboticsReview}.  Collective task allocation decisions require the collective to assign its individual entities to a finite set of tasks.  Collective consensus achievement decisions require the collective to choose one of an available set of options \cite{valentini_achieving_2017}.  Discrete consensus decisions require the collective to select one option from a finite set.  When the highest quality option of a finite set of size $n$ must be chosen, the collective is making a best-of-$n$ decision \cite{Valentini_Survey_2017}.  These decision strategies are often biased towards easily evaluated and discovered options, which reduces decision accuracy when the highest quality option is harder to evaluate than poorer quality options (e.g., \cite{Valentini2014:WeightedVoterModel,Valentini2015:EfficientDecisionMakingSpeedvsAccuracy,ReinaValentini:DesignPattern2015,Valentini_Survey_2017}).  Environmental bias is positive when it increases the collective's ability to choose the highest quality option, but negative when it decreases this ability.   Improving the collective best-of-\textit{n} decision accuracy in the presence of negative environmental bias is an area of continuing research \cite{ReinaValentini:DesignPattern2015,Valentini_Survey_2017,cody_reducing_2017,ReinaICRA:2019}.

Collective decision making changes the relationship in human-collective interaction.  The field of human-swarm interaction has actively pursued the challenges of human control over large groups of robots \cite{HSI_Survey:2016}.  Robotic colonies that effectively determine when the collectives' current activities are completed \cite{Parker2010:UnaryDecisionMaking} or which actions to take next \cite{ReinaValentini:DesignPattern2015} shift the human role in human-collective teams from direct control towards supervision \cite{ParasuramanSheridan:2000}.  Human-collective teams deployed to accomplish environmental monitoring, exploration, or search and rescue tasks are likely to experience communication challenges that reduce the ability of the human to observe and influence the collective.  Enabling collectives to make some decisions autonomously, potentially decreases the interaction requirements of the human operator.  Leveraging collective decision making has been discussed as a means of increasing fault tolerance in human-collective teams by reducing the impact of  human error in decision making processes \cite{HSI_SharedControl}.  Unfortunately, little research exists that examines the sharing of collective decision making between humans and collectives \cite{HSI_SharedControl,HSI_AutonomySpecturmSwarmControl:Coppin2012}. 

This manuscript addresses two challenges necessary to enable the use of human-collective teams: 1) poor collective best-of-$n$ decision accuracy due to environmental bias and 2) the lack of interaction strategies for human-collective decision making. The first challenge is addressed by altering collective best-of-$n$ decisions in order to reduce the influence of negative environmental bias.  The presented experiment requires each collective to choose the highest quality target from a finite set of targets distributed throughout its search range.  Negative environmental bias exists when the highest quality targets are further from a collective's decision making hub than poorer quality targets, and each collective must choose the highest quality target despite these targets taking longer time to discover and evaluate than nearby, poor quality targets \cite{ReinaValentini:DesignPattern2015,ReinaICRA:2019}.   An existing collective best-of-$n$ decision model \cite{ReinaValentini:DesignPattern2015} was combined with collective task sequencing using quorum detection \cite{Parker2010:UnaryDecisionMaking}.  The result is an explicit collective action selection model, analogous to individual action selection (e.g., \cite{Barron:2015}), that enables a collective to choose an option from a set of $n$ options in an environment and execute a task related to that choice \cite{cody_quorum_2017}.  This explicit collective action selection model is compared to a second extended model that includes additional mechanisms that attempt to reduce the influence of environmental bias on collective best-of-\textit{n} decisions \cite{cody_reducing_2017}. The second challenge is addressed by introducing a human-collective interaction strategy that enables the human operator to monitor collectives making best-of-$n$ decisions and issue simple commands in order to influence the decision making process when necessary.  The goal is to enable the humans to better inform collective decisions in human-collective teams and provide the human operator with the means to flexibly influence each collective's autonomy.   

\section{Related Work}
\label{sec:Related_Work}
Collective best-of-$n$ decision strategies have enabled locally sensing groups of robots to determine the best resource (e.g., \cite{CampoGarnierDorigo2011,ReinaValentini:DesignPattern2015,ReinaICRA:2019}), shortest path (e.g., \cite{Schmickl2008,Reina2014,montes_de_oca_majority-rule_2011,DorigoScheidler2016:k-Unanimity}), or future action (e.g., \cite{Parker2010:UnaryDecisionMaking,wessnitzer_collective_2003}).  These decision processes are distinct from collective task allocation strategies, which require the collective to assign individuals to distinct tasks \cite{Brambilla2013:SwarmRoboticsReview}.  A general opinion-based approach to collective best-of-$n$ decision making requires the individual collective entities to maintain opinions favoring one of a finite set of $n$ available options \cite{valentini_achieving_2017}. The collectives' individuals only disseminate their current favoring opinion within the collective decision making hub, in a manner similar to emigration decisions made by \textit{Apis mellifera} honeybee \cite{Seeley2012,Pais2013} and \textit{Temnothorax albipennis} ant \cite{Pratt2001:AntNestSiteDecision} colonies.  The collective individual entities periodically re-evaluate their currently supported option, which may be a specific location (e.g., \cite{Valentini2014:WeightedVoterModel,Valentini2015:EfficientDecisionMakingSpeedvsAccuracy}) or a path (e.g., \cite{montes_de_oca_majority-rule_2011,DorigoScheidler2016:k-Unanimity}), by leaving the hub and performing an assessment.  The dissemination of opinions within the hub is subject to \textit{direct modulation}, as each individual collective entity shares its opinion with a frequency that is proportional to the opinion's value.  Dissemination behaviors are also subject to \textit{indirect modulation} by environmental features that affect option evaluation \cite{valentini_achieving_2017}.  Indirect modulation that increases dissemination frequency for optimal sites introduces a positive environmental bias, while indirect modulation that reduces dissemination frequency for optimal sites introduces a negative environmental bias and often causes poor decisions \cite{Valentini_Survey_2017}.

Individual collective entities alter their opinion based on an individual decision strategy \cite{valentini_achieving_2017}.  Evaluated individual decision strategies include voter models (e.g., \cite{Valentini2014:WeightedVoterModel,Reina2014,Reina2015}), in which individuals randomly select a neighbor's opinion, and majority models (e.g., \cite{montes_de_oca_majority-rule_2011,Valentini2015:EfficientDecisionMakingSpeedvsAccuracy,wessnitzer_collective_2003}), in which the individuals choose the majority opinion within a local neighborhood of individuals. One weakness of these strategies is an inability to choose the best option when that option changes during the decision making process \cite{Dynamic_Decisions_Ferrante_2019}.  Recently, the introduction of a small number of stubborn agents, or the introduction of a small probability of spontaneous opinion switching, have been shown to improve collective decision in dynamic binary decision problems \cite{Dynamic_Decisions_Ferrante_2019}.

One opinion-based decision making strategy that is extended in this manuscript was inspired by \textit{Apis mellifera} honeybee collectives when they choose and emigrate to one of an available set of future locations \cite{ReinaValentini:DesignPattern2015}.  This strategy enables the collective to discover new options during deliberation between options and is value-sensitive, which delays a decision when the choice is between poor quality options \cite{Pais2013,ReinaValentini:DesignPattern2015}.  Reina \textit{et al.} \cite{ReinaValentini:DesignPattern2015} demonstrated that by increasing the time between interactions, the collective was more resilient to negative bias.  Recent efforts have further reduced the influence of negative bias. The use of additional individual agent control states based on distance from the decision making hub \cite{cody_reducing_2017,cody_dissertation_2018} and the modulation of agent interaction rates over time \cite{ReinaICRA:2019} have enabled the collective to make more accurate decisions despite environmental bias.     

This manuscript distinguishes between two types of collective action selection: implicit and explicit.  Most collective best-of-$n$ decision making studies assume that a decision is complete when a consensus, or a majority opinion has been reached \cite{Valentini_Survey_2017}.  The collectives in these scenarios make decisions \textit{implicitly} by selecting an action, such as aggregating in a certain location, or foraging from a certain site, but do not require the collective's entities to respond to the collective's decision \cite{Valentini_Survey_2017}. \textit{Explicit} collective action selection requires the collective to make decisions and alter their behavior once a decision has been made. Altering behavior is common when the collective must determine that a task is completed,  \cite{Parker2010:UnaryDecisionMaking}, or must move to a newly chosen location \cite{cody_quorum_2017}.  

Explicit collective action selection is often achieved through the use of quorum detection mechanisms to halt deliberation during a best-of-$n$ decision process once a majority opinion is established  \cite{ParkerZhang2009:CDMBestOfNProblem}.  When an individual collective entity detected that the number of its peers that shared a similar opinion exceeded a given threshold, it entered into a committed state and transitioned the collective from option deliberation to a decision state \cite{ParkerZhang2009:CDMBestOfNProblem}.  A similar strategy enabled the collective to execute a known sequence of tasks by making a series of decisions about whether the current task was completed \cite{Parker2010:UnaryDecisionMaking}.  A model combining Parker and Zhang's \cite{Parker2010:UnaryDecisionMaking} quorum detection mechanisms with Reina \textit{et al.}'s best-of-$n$ decision strategy enabled a collective to stop deliberation between available options after a decision was reached, without reducing decision accuracy \cite{cody_quorum_2017}.  

Little research examines human interaction with collective task allocation, or collective best-of-$n$ decision making behaviors \cite{HSI_Survey:2016}.  One collaborative task allocation strategy required a team of humans and simulated aerial vehicles to perform a set of tasks \cite{HSI_AutonomySpecturmSwarmControl:Coppin2012}.  The vehicles' actions were determined using a globally shared, virtual pheromone map.  Human operators and robots updated the pheromone map to support information sharing and decision making, which facilitated collaborative task allocation \cite{HSI_AutonomySpecturmSwarmControl:Coppin2012}. A best-of-$n$ shared decision making strategy leveraged Nevai and Passino's \cite{Passino2010} honeybee nest site selection model, in which the collective evaluated and chose the best site from a finite set of $n$ sites. This approach limited human interaction to the collective's decision making hub in an effort to reduce the need for globally shared information \cite{HSI_SharedControl}.

\section{Explicit Collective Action Selection Model}
\label{sec:CAS_Model}
The focal problem starts with the collective best-of-$n$ decision problem, described in Reina \textit{et al.}'s \cite{ReinaValentini:DesignPattern2015} second case study. A collective is required to explore an area and attempt to identify the highest valued site within a search space.  Each site, $i \in \lbrace 1,..,n\rbrace$, had a value, $v_{i} \in(0,1]$ and a location, $loc_{i}\in\mathbb{R}^{2}$ that was a distance, $d_{i}$, from the collective's decision making hub.  An accurate decision occurred when a majority of the individual collective entities favored the highest valued site.  Expanding beyond the original problem, the presented model requires the collective to conduct explicit action selection in order to identify the highest valued site, move the collective's decision-making hub from its original location to the chosen site's location, and initiate a subsequent decision.  All evaluated models use a similar series of quorum detecting states in order to permit the collective to recognize when a decision has been reached and move the collective.  The two evaluated explicit collective action selection models differed according to each model's implementation of the collective best-of-$n$ decision process, which are displayed in Fig. \ref{fig:allModels}.  Individual collective entities interact within the hub, indicated by the white states, and do not interact outside the hub while assessing known sites or exploring, indicated by the shaded states.  

\def \figwidth {0.49}
\begin{figure*}[!t]
    \captionsetup[subfigure]{justification=centering}
	\begin{subfigure}{\figwidth\linewidth}
		\centering
		\includegraphics[keepaspectratio,height=2in]{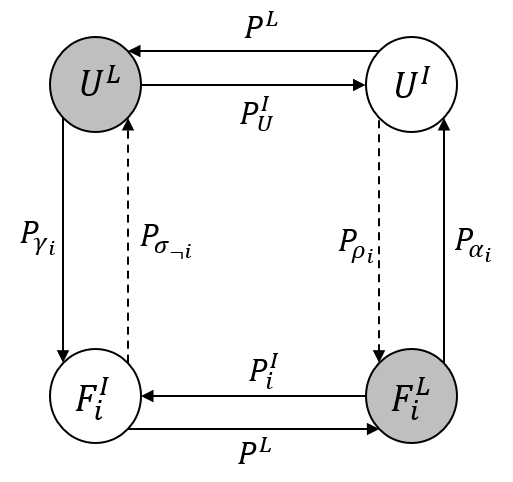}
		\captionsetup{width=\linewidth}
		\caption{$M_{1}$ Best-of-$n$ Model}
		\label{fig:M1_Micro}
	\end{subfigure}
\begin{subfigure}{\figwidth\linewidth}
	\centering
	\includegraphics[keepaspectratio,height=2in]{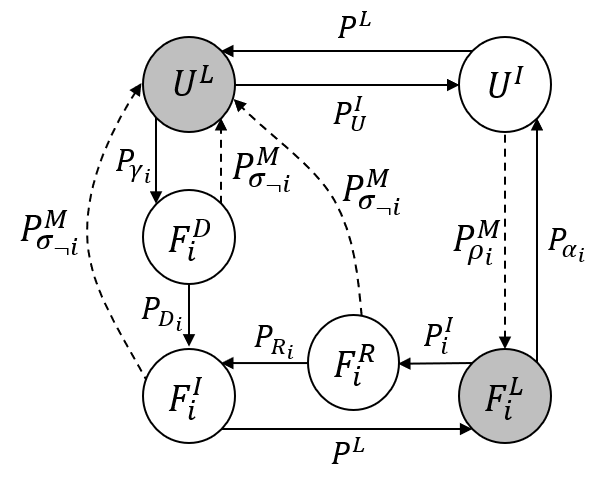}
	\captionsetup{width=\linewidth}
	\caption{$M_{2}$ Bias Reducing Best-of-$n$ Model}
	\label{fig:M2 Microscopic Model}
\end{subfigure}
	\caption{The original best-of-$n$ process \cite{ReinaValentini:DesignPattern2015}($M_{1}$, left) and the bias reducing extension of the model ($M_{2}$,  right).   Individual collective entities in shaded states are latent and outside the hub.  These entities explore ($U_{L}$) or evaluate ($F_{i}^{L}$) sites.  Individual collective entities in white states are interactive inside the hub and may be recruited ($U^{I}$) or inhibited ($F_{i}^{I}$, $F_{i}^{D}$, $F_{i}^{R}$).}
	\label{fig:allModels}
\end{figure*}

The first model ($M_{1}$), as shown in Fig. \ref{fig:M1_Micro}, implemented Reina \textit{et al.}'s  collective best-of-$n$ decision making process \cite{ReinaValentini:DesignPattern2015}.  A population of $N$ individual collective entities either favor a site $i$ ($F_{i}^{*}~\forall i\in \lbrace 1,..,n\rbrace$), or are uncommitted to any site ($U^{*}$). Uncommitted latent individuals ($U^{L}$) explore outside the hub for sites and, with probability, $P_{\gamma_{i}}$, discover site $i$ and return to the hub in an interactive favoring state  ($F_{i}^{I}$).  Uncommitted latent individuals eventually stop exploring, with probability $P_{U}^{I}$, and return to the hub in an uncommitted interactive state ($U^{I}$).  Uncommitted interactive individuals ($U^{I}$) are recruited by interactive favoring individuals ($F_{i}^{I}$), with probability $P_{\rho_{i}}$, and enter the favoring latent state ($F_{i}^{L}$), or resume exploration ($U^{L}$), with probability $P^{L}$.  Favoring latent individuals ($F_{i}^{L}$) assess site $i$ before returning to the hub, but may abandon their site with probability $P_{\alpha_{i}}$ and become uncommitted interactive ($U^{I}$).  Favoring interactive individuals ($F_{i}^{I}$) are subject to being inhibited by individuals supporting sites other than $i$, with probability $P_{\sigma_{\neg i}}$, which causes the inhibited individuals to enter the uncommitted latent state, $U_{L}$.  Individual collective entities conducted correlated random walks \cite{codling_random_2008} while within the dissemination hub and while exploring within the range of the collective, otherwise, the individuals moved directly between known sites and the hub.  The random walks mixed the population within the hub, which frequently altered any one individual's neighbors between that individual's opinion transmissions \cite{ReinaValentini:DesignPattern2015}.

Environmental bias influences the original model ($M_{1}$)  when site value is independent of site location \cite{ReinaValentini:DesignPattern2015,cody_reducing_2017,cody_dissertation_2018}.  Sites that are further away from the decision making hub are less likely to be chosen, because they have a lower probability of discovery ($P_{\gamma_{i}}$) and require more time for individual collective entities in state $F_{i}^{L}$ to evaluate, which decreases $P_{i}^{I}$ \cite{cody_reducing_2017}.  The second model, $M_{2}$ in Fig. \ref{fig:M2 Microscopic Model}, extends model $M_{1}$ with two mechanisms that attempt to compensate for environmental bias: \textit{interaction delay} and \textit{interaction frequency modulation} \cite{cody_dissertation_2018,cody_reducing_2017}. Interaction delay introduces two semi-interactive control states whose exit probabilities are determined by comparing the site's distance from the decision making hub, $d_{i}$, to the minimum and maximum site distances ($d_{min}$ and $d_{max}$) that define the collective's search space.  After site discovery, individual collective entities enter state $F_{i}^{D}$ that they leave with probability $P_{D_{i}}$, which is inversely proportional to the difference between the site's discovery time and the discovery time of a site located at $d_{max}$.  Similarly, individuals that return from site reassessment, enter state $F_{i}^{R}$, where they remain with probability $P_{R_{i}}$, which is inversely proportional to the difference between their current site's round trip time, $\tau_{i}$, and the round trip time to a site at the maximum distance, $\tau_{max}$. Interaction frequency modulation affects inhibition and recruitment rates between individual collective entities.    

The transition probabilities for both models depend on site value, $v_{i}$, and site distance, $d_{i}$.  The interactive transitions are related to the ratio of the individual entities supporting a site, $N_{i}$, to the total number of individual entities in the collective, $N$.  Transition probabilities between latent and interactive states were determined using either the average round trip travel time, $\bar{\tau}$, or the actual round trip travel time to a site, $\tau_{i}$.  The transition probabilities for model $M_{1}$ are described as:
\begin{align}
P_{\alpha_{i}}&=\alpha v_{i}^{-1};\quad P_{\gamma_{i}}=v_{i}g(d_{i});\quad
g(d_{i})=\frac{v_{i}\mu~e^{\xi~d_{i}}}{d_{i}};\quad
P_{\rho_{i}}=\frac{v_{i}N_{i}}{N};\quad P_{\sigma_{\neg i}}=\sum\limits_{j\ne i}\frac{v_{j}N_{j}}{N},\label{eq:fourTransitions}\\
P_{U}^{I}&=\bar{\tau}^{-1};\quad P_{i}^{I}=\tau_{i}^{-1};\quad
P_{L}=9P_{U}^{I},\label{eq:latencyProbs}
\end{align}
where $\alpha$ is the minimum abandonment rate, set to $5$\% per meter traveled \cite{ReinaValentini:DesignPattern2015,cody_reducing_2017}, and $g(d_{i})$ is the estimated discovery probability, where $d_{i}$ is the distance of site $i$ from the collective hub and the values $\mu=0.058$ and $\xi=-0.29$ were determined by fitting a curve to discovery data that was gathered from fifty simulations of the collective's exploration behavior \cite{ReinaValentini:DesignPattern2015,cody_reducing_2017}. Equation \ref{eq:fourTransitions} shows that abandonment probability ($P_{\alpha_{i}}$) increases for lower quality sites and that higher valued sites result in higher discovery ($P_{\gamma_{i}}$), recruitment ($P_{\rho_{i}}$), and inhibition ($P_{\sigma\neg i}$) rates.  The transition probabilities associated with interaction delay and frequency modulation for model $M_{2}$ are described by:

\begin{align}
P_{D_{i}}&=\frac{v_{i}}{g(d_{max})^{-1}-g(d_{i})^{-1}};\quad
P_{R_{i}}=\frac{v_{i}}{\tau_{max}-\tau_{i}},\label{eq:int_delay}\\
M_{i}&=\frac{d_{i}}{d_{min}};\quad P_{\sigma_{\neg i}}^{M}=\sum\limits_{j\ne i}\frac{M_{j}v_{j}N_{j}}{N};\quad P_{\rho_{i}}^{M}=\frac{M_{i}v_{i}N_{i}}{N},\label{eq:interact_eq}
\end{align}
where Eq. \ref{eq:int_delay} defines the exit probabilities of states $F^{D}_{i}$ and $F^{R}_{i}$ as inversely proportional to the difference between a site at $d_{max}$ and site $i$'s discovery and round trip times.  The interaction frequency modulation equations (Eq. \ref{eq:interact_eq}) are defined in terms of $M_{i}$, which is the ratio of the the distance to site $i$ from the decision making hub and the minimum site distance.  As $d_{i}$ increases, the recruitment and inhibition attempt frequency for individual entities supporting site $i$ also increases.  Site distance is assumed to be less than $d_{max}$, which means $g(d_{i}) < g(d_{max})$ and $\tau_{i} < \tau_{max}$.

Both models were extended with a series of quorum detecting states, similar to an existing task sequencing model \cite{Parker2010:UnaryDecisionMaking}, in order to permit the collectives to stop deliberation once a site had been chosen and move to the chosen site.  Each individual collective entity maintained a queue of their last $k=15$ interactions, similar to the original task sequencing model \cite{Parker2010:UnaryDecisionMaking}.  The queues were cleared each time an individual changed states \cite{cody_quorum_2017}.  Individuals detected a quorum after receiving $k$ messages in a row from other individuals in the same state, supporting the same site.  Queue messages were sent with probability $M_{i}P_{U}^{L}$.  Upon detecting a quorum, favoring individuals ($F_{i}^{I}$) entered a committed state ($C_{i}$), remained in the hub, and caused individuals in the deliberation states to commit to site $i$ upon interaction.  Committed individuals transitioned to an action initiation state ($I_{i}$) after detecting a quorum, and interacted with individual collective entities in states $F_{i}^{I}$, $U^{I}$, and $C_{i}$ in order to transition to the initiation state.  Initiating individuals started execution ($X_{i}$) after detecting a quorum, or after waiting $k$ message intervals without receiving a queue message.  Executing individuals moved from the collective's current hub location to the location of the chosen site, $i$, and entered the completed, or done ($D_{i}$) state.  Finally, individual collective entities in state $D_{i}$ transitioned back to the uncommitted interactive state ($U^{I}$) after detecting a quorum and began the next decision from the new location.  

\section{Human-Collective Interface}
\label{sec:HCI_Interface}
Three requirements guided the design of the human-collective interface for the action selection processes described in Section \ref{sec:CAS_Model}: collective state transparency \cite{HSI_SwarmTransparency}, restricted communication with the decision making hub \cite{HSI_SharedControl}, and flexible collective autonomy \cite{HSI_AutonomySpecturmSwarmControl:Coppin2012}.  First, collective state transparency requires that the interface permits rapid observation of the collectives' decision making processes \cite{HSI_SwarmTransparency}.  Second, restricting communication to the collectives' decision making hubs avoids assuming a global communication capability \cite{HSI_SharedControl}. Finally, flexible collective autonomy implies that each collective is capable of making independent decisions \cite{ParasuramanSheridan:2000}, but that the operator adjusts the collective's autonomy using control mechanisms (e.g., the operator allows the collective to make an independent decision or takes over the decision process as needed). Behavior selection controls \cite{HSI_Beacon_Intermittent_Swarm_Influence:Kolling2013,HSI_AutonomySpecturmSwarmControl:Coppin2012}, which permit the operator to transition individual collective entities between internal states, were used to simultaneously alter the collectives' level of autonomy and guide collective decision outcomes.  The remainder of this section describes the interface visualization and the control mechanisms used for influencing the collectives.

\begin{figure}[!t]
	\centering
	\includegraphics[width=\linewidth]{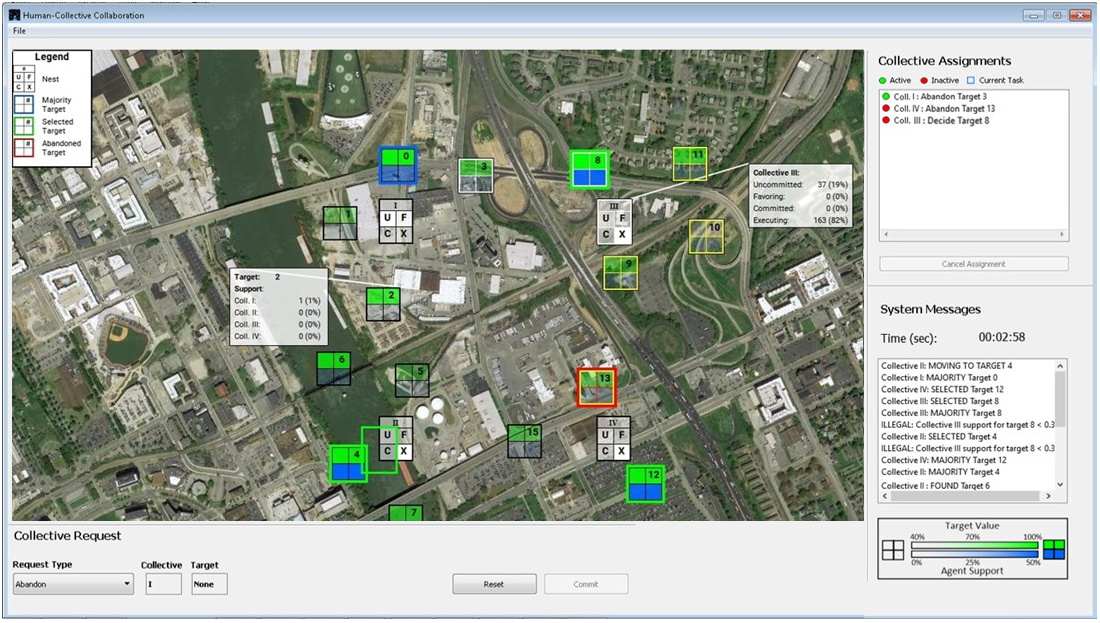}
	\caption[Human-Collective Prototype Interface Screenshot]{The human-collective interface mid-way through a scenario, showing the current locations of the four collectives and the locations of the discovered targets.  Collective decision making hubs are the boxes indicated by Roman numerals. Targets are the green and blue boxes indicated by integer identifiers.  The top half of each target indicates the target's relative value (green) and the bottom have indicates the support of the highest supporting collective (blue).}
	\label{fig:hciscreenshot}
\end{figure}

\subsection{Interface Visualization} 

\label{sec:HCI_CollectiveVisualization}
 
The \textit{Collective Interface} enabled an operator to interact with multiple collectives simultaneously and each collective was required to choose the highest valued target within its search range.  A screenshot of the  \textit{Collective Interface} is provided in Fig. \ref{fig:hciscreenshot}. Collectives and discovered targets were geo-located on the map area in the center of the interface.  The Collective Requests area (lower left) enabled the operator to change individual collective entities' behavior states in a manner explained in Section \ref{sec:HCI_CollectiveControls}.  The Collective Assignments area (upper right) provided a summary of active and inactive requests sent to the respective collectives.  Finally, the System Messages area (lower right) provided text alerts of collective activities (e.g., finding targets) and human input error feedback.  
Four collectives and sixteen targets are shown on the map in Fig. \ref{fig:hciscreenshot}.  

Box symbols represented collectives and targets.  Collectives were white boxes identified by a Roman numeral above four quadrants labelled for the individual collective entity states (as described in Section \ref{sec:CAS_Model}): \textit{uncommitted} ($U$), \textit{favoring} a target ($F$), \textit{committed} to a target ($C$), or \textit{executing} a move to a target ($X$).  The executing state included collective individuals in the previously described initiating ($I$) and done ($D$) states.  The opacity of each quadrant  indicated the percentage of the population in that state. Collective I, in Figure \ref{fig:hciscreenshot}, is primarily uncommitted or favoring, as indicated by the largely opaque $U$ and $F$ quadrants.  Targets were green and blue boxes, labelled in the top right with an integer value.  The opacity of a target's green upper portion indicated the quality of the target, with brighter green indicating higher quality.  The opacity of the blue area of each target indicated the highest percentage of favoring agents supporting that target.  Target 0, above Collective I in Figure \ref{fig:hciscreenshot}, is a high valued target (bright green top) with some support (semi-transparent blue bottom).  Newly discovered targets were initially transparent, but displayed a green color as soon as two individual collective entities favored the target and returned to the hub.  Early in the decision process, target values fluctuated as a result of the collective members' noisy estimates of target value. A target was highlighted with a blue outline when more than $30$\% of a collective's individuals supported it, indicating a collective was about to commit to a target (see Target 0 in Fig. \ref{fig:hciscreenshot}).  The target was highlighted with a green outline as individual collective entities began to execute the movement to a new location (see Target 8 in Fig. \ref{fig:hciscreenshot}).  The Collective II symbol indicates that the collective is executing a movement from its current location to Target 4.  The movement of Collective II is represented by the green box, which moves from the collective's current position to the chosen target's location.  At the end of the movement, the Collective II symbol replaces the Target 4 symbol.

Left clicking on a hub, or a target, selected it as the designated object of the operator's collective request  (see Section \ref{sec:HCI_CollectiveControls}).  Left clicking on a hub also highlighted targets that were in range of the collective and were supported (white outline) or not supported (yellow outline).  Right clicking on targets revealed an estimate of each collectives' support for the target, as shown with Target 2.  Right clicking on a hub revealed a detail flag, as shown for Collective III, with an estimate of the number of collective members in each of the four previously described states.   

The abstract visualization enabled the operator to quickly identify the state of the collective's decision making process.  The operator supervised collective decisions using the Collective Interface.  The operator influenced the decision making process using available behavior selection controls.

\subsection{Collective Controls}
\label{sec:HCI_CollectiveControls}

The operator adjusted the collective's autonomy by completing \textit{investigate}, \textit{abandon}, or \textit{decide} requests using the Collective Request area located on the lower left hand side of Fig. \ref{fig:hciscreenshot}.  Operator activities that were not related to requests were considered observation actions.  Recorded collective observation actions included determining which targets were in range of a particular collective and extra left clicks on targets.  Completed operator requests were added to the Collective Assignments area shown at the top right of Figure \ref{fig:hciscreenshot}.  Erroneous requests were identified in order to prevent the operator from attempting to have collectives investigate targets outside their range, abandon targets that hadn't yet been evaluated, or decide for the collective prior to the collective exploring its search space.  Erroneous requests were ignored and caused an error message to be displayed to the operator in the Systems Messages Area, which was located at the bottom right of the simulation.    

The operator issued an \textit{investigate} request to increase a collective's support for a specific target by transitioning ten uncommitted entities (5\% of the collective population) to the favoring state supporting the chosen target. Additional support for the same target was achieved by reissuing the investigate request. Investigate requests for targets outside the collective's range were unacceptable and reported to the operator as errors. The \textit{abandon} request reduced a collective's support for a specific target by transitioning favoring individual entities in the decision making hub to the uncommitted state. The abandon request only needed to be issued once in order for the collective to ignore a target.  Abandon requests for targets with less than $2$ favoring agents were considered erroneous and reported to the operator.  The \textit{decide} request committed two individual collective entities ($1$\% of the population) to the target designated by the operator.  These individuals then committed other individuals through interactions that rapidly drove the collective to execute a movement to the operator's chosen target.  Issuing a decide request before at least $30$\% of the collective's population favored that target resulted in an error message.   

Issuing a request required the operator to select the desired request type from the drop down menu, select the desired collective and target, and select the commit button to execute the request  (see lower right of Fig. \ref{fig:hciscreenshot}). The reset button cleared the request information without sending the request.  Recently executed requests (e.g., Collective I: Abandon Target 3) were displayed in the upper right hand corner of the monitor area, in the Collective Assignments area. Green and red circles next to each request signified whether the request was in progress (green) or completed (red). Investigate request circles changed from green to red after ten individual entities received and acknowledged the investigate request. Abandon requests remained active (green), once issued. Once a collective reached a decision, all prior requests associated with that particular collective were removed from the collective assignments area. Abandon requests were the only requests the operator was able to cancel, which required selecting the desired request in the Collective Assignments area and pushing the Cancel Assignment button.

\section{Experimental Design}
\label{sec:HumanCollectiveExperimentalDesign}

Two experiments were conducted to test the following hypotheses.  $H_{1}$ predicted that incorporating the bias-reducing mechanisms described in Section \ref{sec:CAS_Model} increases the accuracy of collective best-of-$n$ decisions. $H_{2}$ predicted that human-collective teams outperform the independent collectives in both action selection decision accuracy and decision time.  Finally, $H_{3}$ predicted that the Collective Interface enables the human operator to understand and modify the collective action selection behavior.  The first experiment was simulation based and compared the behavior of the original collective best-of-$n$ decision process, model $M_{1}$, to that of the bias-reducing decision process, model $M_{2}$, without humans.  The second experiment was human-oriented and compared the performance of the two decision process models within human-collective teams.  A third model was added to the second experiment, called $M_{3}$, which served as a control and did not include an independent best-of-$n$ decision process.  Model $M_{3}$ is described in Section \ref{sec:Experimental_HCI_Teams}.

Both experiments involved four  simulated robotic collectives that were required to make targeting decisions within an approximately $2$ km$^{2}$ environment.  Each collective had $200$ simulated Unmanned Aerial Vehicles (UAVs) that were limited to a $500$m exploration radius from the collectives' central decision making hubs.  Simulated robots behaved according to the models described in Section \ref{sec:CAS_Model}, which enabled each collective to perform explicit action selection within its coverage area.

Similar trials were used in both experiments and were modelled after the problem described in Section \ref{sec:CAS_Model}.  At the start of each trial, four collectives were positioned $600$m apart in a square formation, centered within the environment, which purposely overlapped the collectives' search spaces.  Each collective's starting location had four targets that were randomly located at distances within the set  $\lbrace 150, 250, 350\rbrace$m.  The values assigned to the targets were within the set $\lbrace 0.7,0.8,0.9,1.0\rbrace$ and were assessed by individual collective entities with a $10$\% error.  Target values were assigned based on initial decision difficulty.  Easy initial decisions placed the optimal site value ($1.0$) closest to the collective decision making hub (e.g., at distance of $150$m), while difficult initial decisions placed the optimal site value ($1.0$) furthest from the decision making hub (e.g., at a distance of $350$m).  Collectives were visible at the start of the trial and targets were revealed after at least one collective individual discovered it and returned to the collective's hub.  Trials were randomly generated according to these specified site values and distances. 

Each trial was divided into two sections: one with easy initial decisions and one with difficult initial decisions.  Each section required at least six decisions (for a total of at least 12 decisions in the trial).  Following the collective's first target selection and occupation of its chosen target site, the collective's next decision was either easy or difficult, depending upon the collective's location relative to the remaining targets.  The collective hubs were reset to their initial locations before commencing the second trial section, and all targets from the prior trial section were removed. 

A larger set of individual robot capabilities than those typically used in collective best-of-$n$ decision evaluation were assumed (e.g., \cite{ParkerZhang2009:CDMBestOfNProblem,Valentini2014:WeightedVoterModel,ReinaValentini:DesignPattern2015}).  Simple ground platforms, such as Kilobots \cite{kilobot} and e-Puck robots \cite{mondada_e-puck_2009} have been optimized for laboratory research. Both types of robots operate for long periods of time between battery charges (e.g., 2 hours for the e-Puck and several hours for the Kilobot) and are capable of short range ($<25$cm) peer to peer communication.  Each capability is central to many biologically-inspired collective decision making strategies (e.g.,  \cite{ParkerZhang2009:CDMBestOfNProblem,Valentini2014:WeightedVoterModel,ReinaValentini:DesignPattern2015,haque2016}), but are difficult to recreate in UAV collectives.  UAV peer-to-peer communication is complicated by the UAVs' high mobility \cite{FANET_Challenges,FSO_Majumdar2015}.   Many commercial rotor wing UAVs are capable of speeds of $60$km/h, but have limited flight times of approximately $25$ minutes (e.g., \cite{UAVCompare}). Despite these challenges, the simulated agents were assumed to have emergent capabilities and moved at $60$km/h for at least an hour of flight time and communicated via $30$m peer-to-peer transmissions (e.g., optical transmissions \cite{FSO_Alignment}).  A simplifying assumption was that robots moved at constant speed, which is similar to prior research \cite{cody_quorum_2017,ReinaValentini:DesignPattern2015}, but notably deviates from practical UAV movement.

The remainder of this section describes the two experiments in detail.  Section \ref{sec:Experimental_Design_Best_of_N} describes the experiment and metrics used to compare the basic collective action selection model,  Model $M_{1}$, to the bias-reducing model, Model $M_{2}$.  Section \ref{sec:Experimental_HCI_Teams} describes the methods and metrics used to compare Models $M_{1}$, $M_{2}$, and $M_{3}$ as part of human-collective teams in human trials.

\subsection{Experiment 1: Independent Collective Action Selection Models}
\label{sec:Experimental_Design_Best_of_N}

The first experiment tested hypothesis $H_{1}$ by comparing the explicit action selection models described in Section \ref{sec:CAS_Model} in a series of target selection decisions with no human influence.  The primary independent variable was the original ($M_{1}$), or bias-reducing ($M_{2}$) collective action selection models.  Each model was evaluated 10 times for 28 trials, as described at the beginning of this section.  Secondary independent variables included the locations of the targets with respect to each collective's decision making hub, the targets' values, and decision difficulty (e.g., easy or difficult).  Each collective made six decisions in each trial section, for a total of twelve decisions per trial.       

The dependent variables for comparing the models' performance were success rate and decision time.  Success rate was the ratio of the correct number of decisions made by the collective to the total number of decisions.  A collective made correct decisions by identifying and moving to the highest valued target within its search area.  Decision time was the time from the start of a collective decision to the completion of the collective's movement to its chosen target.  The success rate and decision times were averaged over the ten runs of each of the 28 trials in order to retain a similar sample size when comparing to the human performance in the second experiment.  

\subsection{Experiment 2: Collective Action Selection Models in Human-Collective Teams}
\label{sec:Experimental_HCI_Teams}

The second experiment evaluated the two collective action selection models and a baseline model within human-collective teams using the Collective Interface Visualization.  The primary and secondary independent variables were identical to the first experiment; however, a baseline model, called $M_{3}$, was created.  Unlike models $M_{1}$ (original) and $M_{2}$ (bias reduction), $M_{3}$ did not make independent decisions or deliberate between targets (e.g., recruitment and inhibition interactions were disabled). The human operator increased support for targets and made decisions for the $M_{3}$ collective using investigate and decide requests, respectively.  The $M_{3}$ model established operators' baseline performance with a low autonomy collective decision making algorithm. 

Each operator completed three twenty minute trials, with each trial corresponding to one of the collective action selection models.  The trials were assigned such that half the operators completed the evaluation with $M_{1}$ followed by $M_{2}$, and the other half experienced the $M_{2}$ model first.  Additionally, the trials were assigned such that half the operators for each model started the trial with difficult initial decisions and the other half began with easy initial decisions. All operators conducted the baseline $M_{3}$ trial last, with the easy initial decisions preceding the hard decisions.  The baseline model was expected to benefit from the learning effects related to the problem, interface, and collective behavior when compared to the primary models.  Each trial required 12 collective target selections to be made by the four collectives simultaneously, which ensured at least two collectives made two consecutive decisions in each trial section.  

The dependent variables were success rate, decision time, request frequency, and intervention rate.  Success rate and decision time were defined in Section \ref{sec:Experimental_Design_Best_of_N}.  Request frequency was the number of requests, per decision, per minute issued by the human operators during the trials.  The frequency of each request type (e.g., investigate, abandon, and decide) was also gathered.  Intervention rate was the number of interventions for 12 decisions per operator. An intervention occurred when the collective had achieved at least $10$\% support for a target and the operator issued an abandon request.   Interventions indicated that the operator overrode the collective's decision making process in order for the collective to choose a different target.  

Additional objective metrics included the operator actions associated with observing a collective's state (e.g., clicking on a collective or target in order to view state information).  During the trials, operators were required to answer probe questions \cite{Curtis} used to determine their level of situational awareness \cite{SA_Endsley}.  Situational Awareness (SA) probe questions can determine a human operator's situational awareness during critical tasks \cite{Curtis} according to \citeauthor{SA_Endsley}'s \cite{SA_Endsley} three levels of situational awareness: perception, comprehension, and projection.  Perception (Level 1) questions determined the operators' ability to perceive the targets and collectives as well as attributes associated with each (e.g., \enquote{Which collective is investigating Target 1?}).  Comprehension (Level 2) questions determined operators' understanding of perceived elements in relation to collective decisions (e.g., \enquote{Which target is the best choice for Collective III?}).  Finally, projection (Level 3) questions determined operators' ability to estimate the collective's future state based on their perception and comprehension of the current state (e.g., \enquote{Which collective will make the next decision?}).  Operators were asked four Level 1, five Level 2, and three Level 3 SA probes for each model.  The SA probes were similar across operators, but were written as templates and the specific collective or target was added during the experiment.  The Level 1 question, \enquote{Which Collective is investigating Target \_?}, for example, was completed with an applicable target number during the trial, before asking the SA probe question.  

Recorded subjective data included answers to a demographic questionnaire, performance on  \citeauthor{MRT_vandenberg_mental_1978}'s \cite{MRT_vandenberg_mental_1978} Mental Rotations Test (MRT), and a post experiment questionnaire that rank ordered the models. Each trial ended with a post trial questionnaire, the NASA Task Load Index (NASA-TLX) and a 3-D Situation Awareness Rating Technique (3-D SART)   \cite{SART_selcon1991workload}.  The NASA-TLX provided a workload estimate, which included the weighted summation of a variety of workload components including mental demand, effort, and frustration. The 3-D SART score for situational awareness was calculated using the perceived Situational Understanding (SU), Demands on Attentional Resources (DAR), and Supply of Attentional Resources (SAR), according to the following equation: SART Score = SU - (DAR - SAR).  The post-trial questionnaires focused on the perceived performance and the responsiveness of the collective during the trial.  The post experiment questionnaire required the operators to rank order the different collective models according to responsiveness, performance, and ease of comprehension.  

Twenty-eight operators from the Vanderbilt University campus and surrounding area completed the experiment.  The 15 female and 13 male operators were predominantly in the 18 to 30 year age range, although four operators were between the ages of 31 and 50.  The operators had at least completed high school.  More than half the operators had completed (13 operators), or were completing (11 operators) an undergraduate degree. Each operator began the experiment by completing the informed consent paperwork, the demographic questionnaire, and the MRT.  Once these items were completed, the operators received a scripted introduction to the experiment and the simulator ($5$ minutes).  Prior to each trial, the operators conducted $5$ minute training sessions with the specific model that consisted of two collectives, with one collective required to make an easy decision and one required to make a difficult decision.  Operators responded to the SA probes at increments of approximately one probe per minute during each trial.  At the end each trial, the operators completed the post-trial questionnaire, the NASA-TLX, and the 3-D SART.  After all trials and post-trial data collection, the operators completed the post-experiment questionnaire.

\section{Results}
\label{sec:OverallResults}
\makeatletter
\newcommand{\thickhline}{%
    \noalign {\ifnum 0=`}\fi \hrule height 1pt
    \futurelet \reserved@a \@xhline
}
\newcolumntype{"}{@{\hskip\tabcolsep\vrule width 1pt\hskip\tabcolsep}}
\makeatother

\begin{table}[bp!]
\centering
	\captionsetup{aboveskip=3pt}
	\caption[Independent Collective Results:]{Success Rate (\%) Per Decision.}
\begin{tabular}{c|c|c"c|c"c|c|}
\cline{2-7}
& \multicolumn{2}{c"}{Overall} & \multicolumn{2}{c"}{Easy} & \multicolumn{2}{c|}{Difficult} \\ 
\cline{2-7}
& \cellcolor{Gray}Mean & \cellcolor{Gray}Median & \cellcolor{Gray}Mean & \cellcolor{Gray}Median & \cellcolor{Gray}Mean & \cellcolor{Gray}Median \\
& (SD) & (Min/Max) & (SD) & (Min/Max) & (SD) & (Min/Max) \\ \thickhline
\multicolumn{1}{|c|}{\multirow{2}{*}{$M_{1}$ SIM}} & \cellcolor{Gray}52.11 & \cellcolor{Gray}60 & \cellcolor{Gray}70.49 & \cellcolor{Gray}87.5 & \cellcolor{Gray}6.8 & \cellcolor{Gray}0 \\
\multicolumn{1}{|c|}{} & (39.41) & (0/100) & (35.29) & (0/100) & (17.85) & (0/100) \\ \thickhline
\multicolumn{1}{|c|}{\multirow{2}{*}{$M_{2}$ SIM}} & \cellcolor{Gray}74.58 & \cellcolor{Gray} 70 & \cellcolor{Gray}76.7 & \cellcolor{Gray} 83.33 & \cellcolor{Gray}64.04 & \cellcolor{Gray} 66.67 \\
\multicolumn{1}{|c|}{} & (18.39) & (20/100) & (26.87) & (0/100) & (26.38) & (0/100) \\ \hline
\end{tabular}
\label{table:independent_SR}
\end{table}

\begin{table}[bp!]
\centering
	\captionsetup{aboveskip=3pt}
	\caption[Independent Collective Results:]{Decision Time (minutes) Per Decision.}
\begin{tabular}{c|c|c"c|c"c|c|}
\cline{2-7}
& \multicolumn{2}{c"}{Overall} & \multicolumn{2}{c"}{Easy} & \multicolumn{2}{c|}{Difficult} \\ 
\cline{2-7}
& \cellcolor{Gray}Mean & \cellcolor{Gray}Median & \cellcolor{Gray}Mean & \cellcolor{Gray}Median & \cellcolor{Gray}Mean & \cellcolor{Gray}Median \\
& (SD) & (Min/Max) & (SD) & (Min/Max) & (SD) & (Min/Max) \\ \thickhline
\multicolumn{1}{|c|}{\multirow{2}{*}{$M_{1}$ SIM}} & \cellcolor{Gray}2.46 & \cellcolor{Gray}2.44 & \cellcolor{Gray}2.18 & \cellcolor{Gray}2.14 & \cellcolor{Gray}3.2 & \cellcolor{Gray}3.05 \\
\multicolumn{1}{|c|}{} & (0.66) & (1.36/4.46) & (0.53) & (1.36/3.89) & (0.96) & (1.83/7.08) \\ \thickhline
\multicolumn{1}{|c|}{\multirow{2}{*}{$M_{2}$ SIM}} & \cellcolor{Gray}4.79 & \cellcolor{Gray} 4.79 & \cellcolor{Gray}4.17 & \cellcolor{Gray} 4.1 & \cellcolor{Gray}5.77 & \cellcolor{Gray} 5.62 \\
\multicolumn{1}{|c|}{} & (1.11) & (2.49/7.7) & (0.93) & (2.49/7.55) & (1.38) & (3.67/10.25) \\ \hline
\end{tabular}
\label{table:independent_DT}
\end{table}

The results of the independent model experiment are presented first and compared to the results of the human-collective team experiments in order to examine the operator's influence on each model's performance.  Section \ref{sec:IndependentCollectiveResults} describes the performance of the collectives in the absence of human influence in terms of success rate and decision time.  Section \ref{sec:HumanCollectiveResults} presents the performance of the human-collective teams, compares the performance of the human-collective teams to the independent model collective experiment, analyzes the actions and awareness of the operators, and presents the subjective results from the post trial and post experimental questionnaires, the NASA-TLX and 3-D SART worksheets, and the MRT scores.

\subsection {Experiment 1: Independent Collective Action Selection Models}
\label{sec:IndependentCollectiveResults}

The descriptive statistics for the independent collectives' performance according to success rate and decision time are provided in Tables \ref{table:independent_SR} and \ref{table:independent_DT}, respectively.  The $SIM$ designation (e.g., $M_{1} SIM$ and $M_{2} SIM$) is used to distinguish the independent model results, from the human-collective teaming results that use the same models.  The bias reducing model, $M_{2} SIM$, achieved higher accuracy with slower decision times, when compared to the original model, $M_{1} SIM$, especially during difficult decisions.  A Mann-Whitney-Wilcoxon test indicated significant effects between the models for overall success rate, \textit{N = 672, degrees of freedom (DOF) = 1, U = 41017, $\rho$ $<$ 0.001}, and success rate for difficult decisions,  \textit{N = 456, DOF = 1, U = 3157.5, $\rho$ $<$ 0.001}.  A Mann-Whitney-Wilcoxon test also indicated significant effects between the models for decision time: overall decisions - \textit{N = 672, DOF = 1, U = 4320.5, $\rho$ $<$ 0.001}, easy decisions - \textit{N = 578, DOF = 1, U = 1851, $\rho$ $<$ 0.001}, and difficult decisions - \textit{N = 456, DOF = 1, U = 2617, $\rho$ $<$ 0.001}.   

The results indicate that model $M_{1} SIM$ selected the best site consistently when that site was closer to the collective's decision making hub than other poorer quality sites.  During difficult decisions, $M_{1} SIM$ selected a nearby poor quality site rapidly, even after discovering the best site.  The bias reducing model, $M_{2} SIM$, achieved a 57\% higher accuracy than $M_{1} SIM$ for the difficult decisions, demonstrating significant resilience to negative environmental bias; however, the model was generally two minutes slower than $M_{1} SIM$ during easy decisions.       

Further analysis revealed a strong negative correlation between decision time and success rate when using $M_{1} SIM$ for all decisions, \textit{r = -0.54, $\rho$ $<$ 0.001}, a moderately negative correlation for easy decisions, \textit{r = -0.37, $\rho$ $<$ 0.001}, and a moderately positive correlation for difficult decisions, \textit{r = 0.44, $\rho$ $<$ 0.001}.  A strong negative correlation between decision time and success rate was also observed for $M_{2} SIM$ for all decisions, \textit{r = -0.57, $\rho$ $<$ 0.001}, and a moderately negative correlation for easy decisions, \textit{r = -0.44, $\rho$ $<$ 0.001}.  These correlations suggest that longer decision times tended to result in poor easy decisions by both models.  During difficult decisions, the basic model, $M_{1} SIM$ made just under 7\% correct decisions, but was more likely to be correct when the decision took longer.         

\subsection{Experiment 2: Collective Action Selection Models in Human-Collective Teams}

\label{sec:HumanCollectiveResults}

The results of the Human-Collective Team experiment are presented in four sections.  Section \ref{sec:HumanCollective_Independent_Comparison} compares the success rates and decision times achieved between the performance of the independent collectives ($M_{1} SIM$ and $M_{2} SIM$) and the human-collective teams ($M_{1}$, $M_{2}$, and $M_{3}$).  Section \ref{sec:Human_Collective_Actions} presents the observed human actions during the experiment with models $M_{1}$, $M_{2}$ and the baseline model, $M_{3}$. Section \ref{sec:SA_HC_Comparison} presents the observation actions and responses to the Situational Awareness Probe questions.  Finally, Section \ref{sec:Subjective_HC_Comparison}, presents the subjective results.  

\subsubsection{Comparison Between the Independent Collectives and Human-Collective Team Experiments}
\label{sec:HumanCollective_Independent_Comparison}

\begin{table}[bp!]
\centering
	\captionsetup{aboveskip=3pt}
	\caption[Human Trials Objective Results: Success Rate Performance]{Success Rate (\%) Per Decision.}
\begin{tabular}{c|c|c"c|c"c|c|}
\cline{2-7}
& \multicolumn{2}{c"}{Overall} & \multicolumn{2}{c"}{Easy} & \multicolumn{2}{c|}{Difficult} \\ \cline{2-7}
& \cellcolor{Gray}Mean & \cellcolor{Gray}Median & \cellcolor{Gray}Mean & \cellcolor{Gray}Median & \cellcolor{Gray}Mean & \cellcolor{Gray}Median \\
& (SD) & (Min/Max) & (SD) & (Min/Max) & (SD) & (Min/Max) \\ \thickhline
\multicolumn{1}{|c|}{\multirow{2}{*}{$M_{1}$}} & \cellcolor{Gray}78.57 & \cellcolor{Gray}100 & \cellcolor{Gray}95.31 & \cellcolor{Gray}100 & \cellcolor{Gray}56.25 & \cellcolor{Gray}100 \\
\multicolumn{1}{|c|}{} & (41.09) & (0/100) & (21.19) & (0/100) & (49.78) & (0/100) \\ \thickhline
\multicolumn{1}{|c|}{\multirow{2}{*}{$M_{2}$}} & \cellcolor{Gray}88.39 & \cellcolor{Gray}100 & \cellcolor{Gray}94.44 & \cellcolor{Gray}100 & \cellcolor{Gray}81.41 & \cellcolor{Gray}100 \\
\multicolumn{1}{|c|}{} & (32.08) & (0/100) & (22.97) & (0/100) & (39.03) & (0/100) \\ \thickhline
\multicolumn{1}{|c|}{\multirow{2}{*}{$M_{3}$}} & \cellcolor{Gray}92.86 & \cellcolor{Gray}100 & \cellcolor{Gray}95.94 & \cellcolor{Gray}100 & \cellcolor{Gray}88.49 & \cellcolor{Gray}100 \\
\multicolumn{1}{|c|}{} & (25.79) & (0/100) & (19.79) & (0/100) & (32.03) & (0/100) \\ \hline
\end{tabular}
\label{table:HC_SR}
\end{table}

\begin{table}[bp!]
\centering
	\captionsetup{aboveskip=3pt}
	\caption[Human Trials Objective Results: Decision Time Performance]{Decision Time (minutes) Per Decision.}
\begin{tabular}{c|c|c"c|c"c|c|}
\cline{2-7}
& \multicolumn{2}{c"}{Overall} & \multicolumn{2}{c"}{Easy} & \multicolumn{2}{c|}{Difficult} \\ \cline{2-7}
& \cellcolor{Gray}Mean & \cellcolor{Gray}Median & \cellcolor{Gray}Mean & \cellcolor{Gray}Median & \cellcolor{Gray}Mean & \cellcolor{Gray}Median \\
& (SD) & (Min/Max) & (SD) & (Min/Max) & (SD) & (Min/Max) \\ \thickhline
\multicolumn{1}{|c|}{\multirow{2}{*}{$M_{1}$}} & \cellcolor{Gray}3.01 & \cellcolor{Gray}2.48 & \cellcolor{Gray}2.1 & \cellcolor{Gray}1.88 & \cellcolor{Gray}4.22 & \cellcolor{Gray}4.03 \\
\multicolumn{1}{|c|}{} & (1.56) & (1.16/8.58) & (0.75) & (1.16/4.79) & (1.54) & (1.6/8.58) \\ \thickhline
\multicolumn{1}{|c|}{\multirow{2}{*}{$M_{2}$}} & \cellcolor{Gray}3.97 & \cellcolor{Gray}3.64 & \cellcolor{Gray}3.37 & \cellcolor{Gray}3.09 & \cellcolor{Gray}4.67 & \cellcolor{Gray}4.57 \\
\multicolumn{1}{|c|}{} & (1.37) & (1.83/9.94) & (1.23) & (1.83/9.94) & (1.2) & (2.46/8.81) \\ \thickhline
\multicolumn{1}{|c|}{\multirow{2}{*}{$M_{3}$}} & \cellcolor{Gray}5.32 & \cellcolor{Gray}4.78 & \cellcolor{Gray}4.67 & \cellcolor{Gray} 4.2 & \cellcolor{Gray}6.24 & \cellcolor{Gray}6.03 \\
\multicolumn{1}{|c|}{} & (2.22) & (1.48/13.25) & (1.96) & (1.48/12.78) & (2.24) & (2.07/13.25) \\ \hline
\end{tabular}
\label{table:HC_DT}
\end{table}

The success rate and decision time descriptive statistics for the human-collective teams are provided in Tables \ref{table:HC_SR} and \ref{table:HC_DT} \cite{Roundtree20191}. A Kruskal-Wallis test across models $M_{1} SIM$, $M_{2} SIM$, $M_{1}$, $M_{2}$, and $M_{3}$ identified significant effects for success rate in all decisions, \textit{$\chi^{2}$(4, N = 1680) = 523.39, $\rho$ $<$ 0.001}, easy decisions, \textit{$\chi^{2}$(4, N = 1147) = 381.3, $\rho$ $<$ 0.001}, and difficult decisions \textit{$\chi^{2}$(4, N = 895) = 388.94, $\rho$ $<$ 0.001}.  Significant effects for decision times were observed in all decisions, \textit{$\chi^{2}$(4, N = 1680) = 687.89, $\rho$ $<$ 0.001}, easy decisions, \textit{$\chi^{2}$(4, N = 1147) = 683.52, $\rho$ $<$ 0.001}, and difficult decisions \textit{$\chi^{2}$(4, N = 895) = 376.4, $\rho$ $<$ 0.001}.        

\def \figwidth {0.49}
\begin{figure*}[t!]
    \captionsetup[subfigure]{justification=centering}
	\begin{subfigure}{\figwidth\linewidth}
		\centering
		\includegraphics[keepaspectratio,height=2.5in]{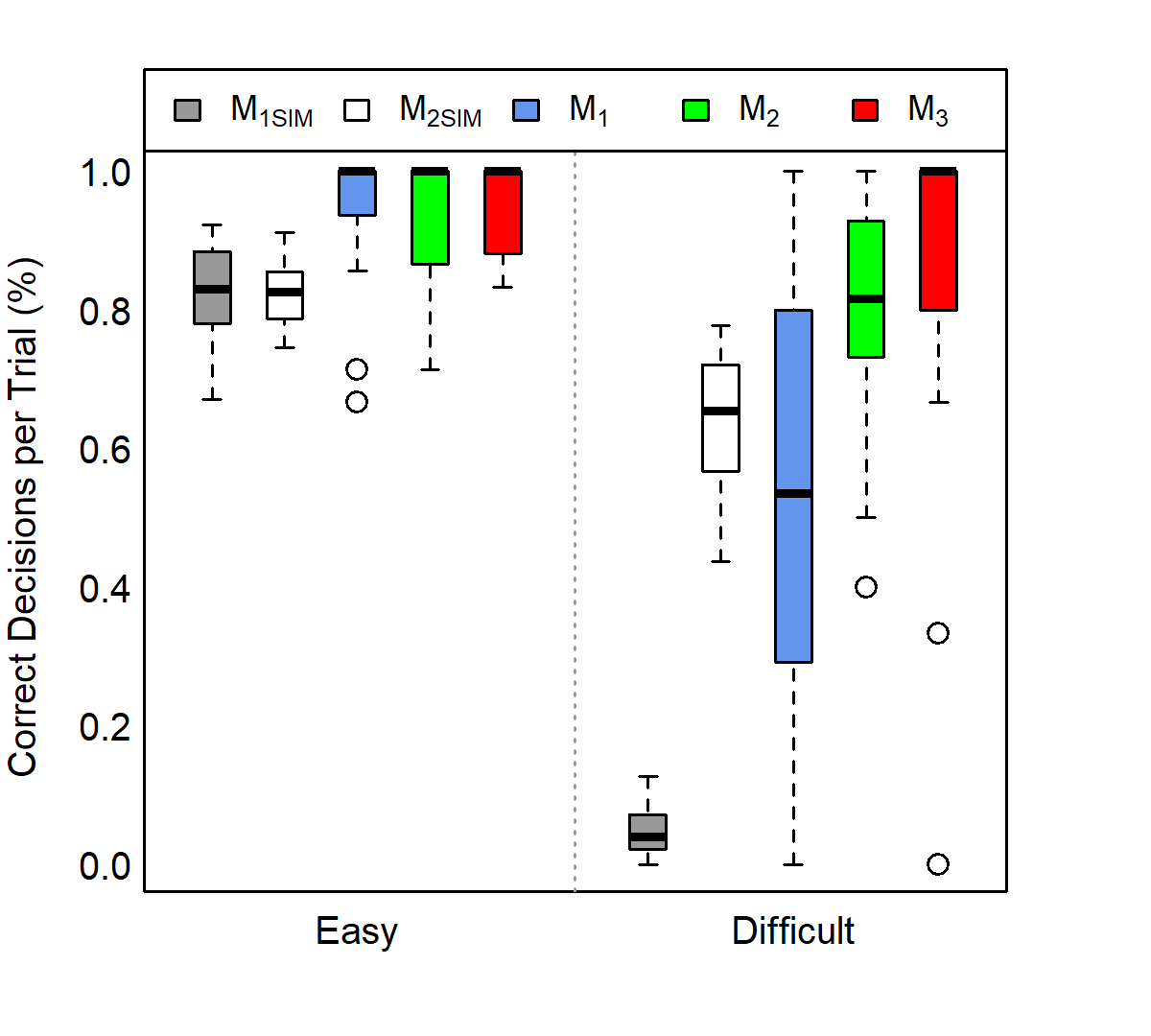}
		\captionsetup{width=\linewidth}
		\caption{Success Rate Comparison}
		\label{fig:Compare_SR}
	\end{subfigure}
\begin{subfigure}{\figwidth\linewidth}
	\centering
	\includegraphics[keepaspectratio,height=2.5in]{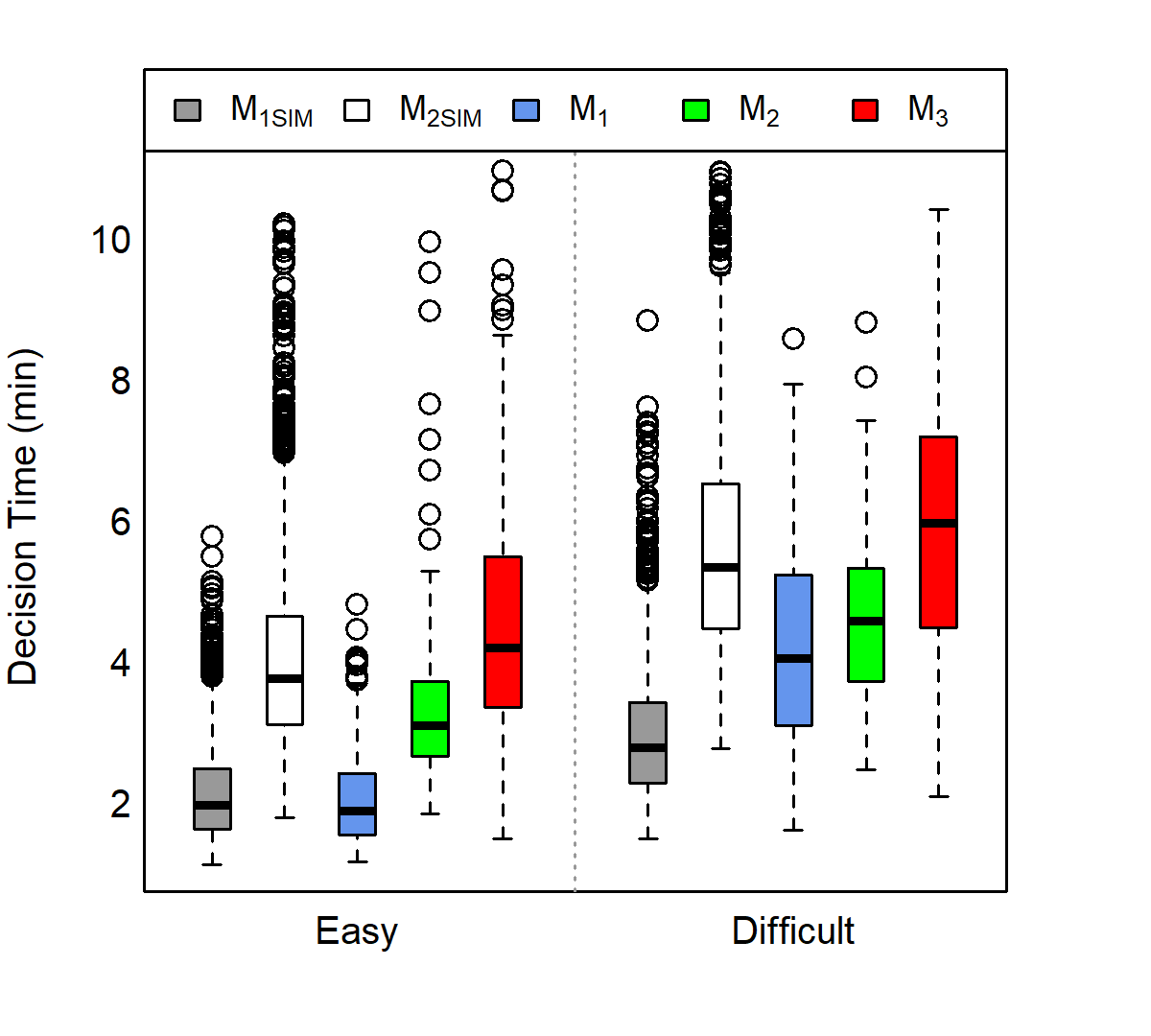}
	\captionsetup{width=\linewidth}
	\caption{Decision Time Comparison}
	\label{fig:Compare_DT}
\end{subfigure}
	\caption{A comparison of the Success Rate and Decision times for the independent collectives, $M_{1} SIM$ and $M_{2} SIM$, and the human-collective teams, $M_{1}$ (original), $M_{2}$ (bias reducing), and $M_{3}$ (baseline).}
	\label{fig:Compare_DT_and_SR}
\end{figure*}

Pairwise comparisons between the independent collectives and human-collective teams indicated significant human influence.  A Tukey and Kramer test revealed significant effects between $M_{1} SIM$ and $M_{1}$ for success rates in overall decisions (\textit{$\rho$ $<$ 0.001}), easy decisions (\textit{$\rho$ $<$ 0.001}), and difficult decisions (\textit{$\rho$ $<$ 0.001}).  A similar test also revealed significant effects between $M_{1} SIM$ and $M_{1}$ for decision times in overall decisions (\textit{$\rho$ $<$ 0.001}) and difficult decisions (\textit{$\rho$ $<$ 0.001}), but not for easy decisions.  The pairwise comparison between $M_{2} SIM$ and $M_{2}$ identified significant effects for success rates in overall decisions (\textit{$\rho$ $<$ 0.001}), easy decisions (\textit{$\rho$ $<$ 0.001}), and difficult decisions (\textit{$\rho$ $<$ 0.001}).  Differences in the decision times between $M_{2} SIM$ and $M_{2}$ were also significant in overall decisions (\textit{$\rho$ $<$ 0.001}), easy decisions (\textit{$\rho$ $<$ 0.001}), and difficult decisions (\textit{$\rho$ $<$ 0.001}).     

The operator's influence is evident in Figure \ref{fig:Compare_DT_and_SR}, which compares the success rates and decision times of the independent models ($M_{1} SIM$ and $M_{2} SIM$) to the operators teamed with the same models ($M_{1}$ and $M_{2}$) and the baseline, human-only, model ($M_{3}$).  Operators improved the success rates of both collective models, although success rates were clearly higher for $M_{2}$ than $M_{1}$ when making difficult decisions, as shown on the right side of  Figure \ref{fig:Compare_DT_and_SR} (\subref{fig:Compare_SR}). Human influence enabled a success rate more than ten times higher than the independent collective, $M_{1} SIM$, for difficult decisions.  Operators decreased decision times for both collective action selection models for easy decisions, but the higher difficult decision accuracy achieved by $M_{1}$ also required the original model to be slowed down, as shown on the right hand side of Figure \ref{fig:Compare_DT_and_SR} (\subref{fig:Compare_DT}).  The operators using $M_{2}$ increased success rates while decreasing decision times in all decisions.  

 Pairwise comparisons also indicated that the collective action selection model significantly affected human-collective team performance.  A Tukey and Kramer test revealed significant effects between the $M_{1}$ and $M_{2}$ for success rates in overall decisions (\textit{$\rho$ $=$ 0.04}) and difficult decisions (\textit{$\rho$ $<$ 0.001}).  Significant effects were also observed between $M_{1}$ and $M_{3}$ for success rates in overall decisions (\textit{$\rho$ $<$ 0.001}) and difficult decisions (\textit{$\rho$ $<$ 0.001}).  No significant effects were observed for success rates between $M_{2}$ and $M_{3}$ for any decision category.  A Tukey and Kramer test revealed significant effects for decision times between all human-collective results ($M_{1}$,$M_{2}$,$M_{3}$) in overall and easy decisions (\textit{$\rho$ $<$ 0.001} for each comparison).  Significant effects were observed in pairwise comparisons between each of the collective action selection models ($M_{1}$ and $M_{2}$) and $M_{3}$ for decision times in difficult decisions (\textit{$\rho$ $<$ 0.001}), but no significant effects were observed for the same metric between $M_{1}$ and $M_{2}$ in difficult decisions.    

The teams with the bias reducing model, $M_{2}$, achieved approximately 9\% higher accuracy overall and 25\% higher accuracy during difficult decisions when compared to the $M_{1}$ teams.  The $M_{2}$ teams required more than a minute longer for each easy decision than the $M_{1}$ teams, although the gap in decision times for difficult decisions was noticeably smaller (less than 30 seconds on average). A moderate positive correlation was found between the decision time and success rates for the $M_{1}$ teams when making difficult decisions, \textit{r = 0.51, $\rho$ $<$ 0.001}.  Weak correlations were observed for the $M_{1}$ teams in overall decisions, \textit{r = -0.19, $\rho$ $<$ 0.001}, for the $M_{2}$ teams in overall, \textit{r = -0.11, $\rho$ = 0.05}, easy, \textit{r = -0.18, $\rho$ = 0.02}, and difficult decisions, \textit{r = 0.18, $\rho$ = 0.03}, as well as for the $M_{3}$ teams in difficult decisions, \textit{r = 0.25, $\rho$ $<$ 0.01}. The analysis suggests that the human-collective teams improved difficult decision accuracy with longer decision times and that this relationship was most common in the $M_{1}$ teams during difficult decisions.  The operators made the fastest decisions using $M_{1}$, but achieved the lowest accuracy in difficult decisions even after slowing the collective decision process down.  The operators using the baseline model, $M_{3}$, were the most accurate, but also the slowest.   Finally, the $M_{2}$ human-collective teams made more accurate decisions than the $M_{1}$ teams and made these decisions faster than operators with the $M_{3}$ teams.

\subsubsection{Human Actions in the Human-Collective Team Experiment}
\label{sec:Human_Collective_Actions}

The significant influence of operators, reported in Section \ref{sec:HumanCollective_Independent_Comparison}, resulted from the actions and awareness of the operators themselves.  This activity was critical to the success of the baseline model, $M_{3}$, which required the human-operators to control the collectives without the aid of a collective decision making process.  The results provided in Tables \ref{table:HC_SR} and \ref{table:HC_DT} as well as in Figure \ref{fig:Compare_DT_and_SR}, show that the human-operators achieved the highest success rates with the baseline model, although these rates were not found to be significantly greater than those of the $M_{2}$ human-collective team.  The baseline decision times; however, were significantly longer than the other two models for all decisions. 

\begin{table}[bp!]
\centering
	\captionsetup{aboveskip=3pt}
	\caption[Human Trials Objective Results: Request Frequency Performance]{Request Frequency (Number of Requests Per Decision/Decision Time).}
\begin{tabular}{c|c|c"c|c"c|c|}
\cline{2-7}
& \multicolumn{2}{c"}{Overall} & \multicolumn{2}{c"}{Easy} & \multicolumn{2}{c|}{Difficult} \\ \cline{2-7}
& \cellcolor{Gray}Mean & \cellcolor{Gray}Median & \cellcolor{Gray}Mean & \cellcolor{Gray}Median & \cellcolor{Gray}Mean & \cellcolor{Gray}Median \\
& (SD) & (Min/Max) & (SD) & (Min/Max) & (SD) & (Min/Max) \\ \thickhline
\multicolumn{1}{|c|}{\multirow{2}{*}{$M_{1}$}} & \cellcolor{Gray}0.57 & \cellcolor{Gray}0.48 & \cellcolor{Gray}0.61 & \cellcolor{Gray}0.54 & \cellcolor{Gray}0.5 & \cellcolor{Gray}0.4 \\
\multicolumn{1}{|c|}{} & (0.56) & (0/2.56) & (0.61) & (0/2.56) & (0.47) & (0/1.98) \\ \thickhline
\multicolumn{1}{|c|}{\multirow{2}{*}{$M_{2}$}} & \cellcolor{Gray}0.66 & \cellcolor{Gray}0.55 & \cellcolor{Gray}0.7 & \cellcolor{Gray}0.6 & \cellcolor{Gray}0.6 & \cellcolor{Gray}0.51 \\
\multicolumn{1}{|c|}{} & (0.49) & (0/2.6) & (0.56) & (0/2.6) & (0.4) & (0/2.03) \\ \thickhline
\multicolumn{1}{|c|}{\multirow{2}{*}{$M_{3}$}} & \cellcolor{Gray}1.33 & \cellcolor{Gray}1.18 & \cellcolor{Gray}1.4 & \cellcolor{Gray}1.21 & \cellcolor{Gray}1.22 & \cellcolor{Gray}1.08 \\
\multicolumn{1}{|c|}{} & (0.79) & (0.11/4.75) & (0.87) & (0.11/4.75) & (0.65) & (0.3/3.66) \\ \hline
\end{tabular}
\label{table:HC_Requests_Per_Decision_Time}
\end{table}

The operator influenced the collectives' behavior directly by issuing the \textit{investigate}, \textit{abandon}, and \textit{decide} requests described in Section \ref{sec:HCI_CollectiveControls}.  The descriptive statistics for the frequency of all requests  issued per decision per minute are summarized in Table \ref{table:HC_Requests_Per_Decision_Time}. The operators issued requests least frequently with $M_{1}$ and slightly more often with the bias reducing model, $M_{2}$.  Requests were most frequent with the baseline model, $M_{3}$, which required consistent human control. A Kruskal-Wallis test revealed significant effects between the three models for request frequency for overall, \textit{$\chi^{2}$ (2, N = 1008) = 239.65, $\rho$ $<$ 0.001}, easy, \textit{$\chi^{2}$ (2, N = 569) = 121.84, $\rho$ $<$ 0.001}, and difficult decisions, \textit{$\chi^{2}$ (2, N = 439) = 120.3, $\rho$ $<$ 0.001}.  A positive correlation was found between request frequency and success rate with $M_{1}$ for all decisions, \textit{r = 0.35, $\rho$ $<$ 0.001}, but the correlation was strongest for the difficult decisions, \textit{r = 0.58, $\rho$ $<$ 0.001}.  A weak negative correlation was found for $M_{3}$ during difficult decisions, \textit{r = -0.19, $\rho$ = 0.03}.  These results suggest that $M_{1}$ was more likely to make accurate decisions when the operator issued additional requests to the collectives, especially during difficult decisions.  Differences in operator request frequencies did not correspond to changes in success rates for the bias reducing model, $M_{2}$, or the baseline model, $M_{3}$.  The $M_{3}$ teams achieved the highest success rates, but required twice as many requests and at least a minute longer per decision than either the basic or bias-reducing collective action selection models. 

\begin{table}[bp!]
\centering
	\captionsetup{aboveskip=3pt}
	\caption[Human Trials Objective Results: Number of Investigate Requests Per Decision Per Minute Performance]{Number of Investigate Requests Per Decision Per Minute.}
\begin{tabular}{c|c|c"c|c"c|c|}
\cline{2-7}
& \multicolumn{2}{c"}{Overall} & \multicolumn{2}{c"}{Easy} & \multicolumn{2}{c|}{Difficult} \\ \cline{2-7}
& \cellcolor{Gray}Mean & \cellcolor{Gray}Median & \cellcolor{Gray}Mean & \cellcolor{Gray}Median & \cellcolor{Gray}Mean & \cellcolor{Gray}Median \\
& (SD) & (Min/Max) & (SD) & (Min/Max) & (SD) & (Min/Max) \\ \thickhline
\multicolumn{1}{|c|}{\multirow{2}{*}{$M_{1}$}} & \cellcolor{Gray}0.44 & \cellcolor{Gray}0.32 & \cellcolor{Gray}0.47 & \cellcolor{Gray}0.43 & \cellcolor{Gray}0.38 & \cellcolor{Gray}0.29 \\
\multicolumn{1}{|c|}{} & (0.49) & (0/2.56) & (0.54) & (0/2.56) & (0.42) & (0/1.65) \\ \thickhline
\multicolumn{1}{|c|}{\multirow{2}{*}{$M_{2}$}} & \cellcolor{Gray}0.47 & \cellcolor{Gray}0.35 & \cellcolor{Gray}0.49 & \cellcolor{Gray}0.34 & \cellcolor{Gray}0.46 & \cellcolor{Gray}0.37 \\
\multicolumn{1}{|c|}{} & (0.44) & (0/2.27) & (0.5) & (0/2.27) & (0.37) & (0/1.63) \\ \thickhline
\multicolumn{1}{|c|}{\multirow{2}{*}{$M_{3}$}} & \cellcolor{Gray}1.07 & \cellcolor{Gray}0.95 & \cellcolor{Gray}1.12 & \cellcolor{Gray}0.97 & \cellcolor{Gray}1 & \cellcolor{Gray}0.92 \\
\multicolumn{1}{|c|}{} & (0.71) & (0/4.3) & (0.79) & (0/4.3) & (0.58) & (0/2.81) \\ \hline
\end{tabular}
\label{table:HC_Investigate_Per_Dec_Time}
\end{table}

\begin{table}[bp!]
\centering
	\captionsetup{aboveskip=3pt}
	\caption[Human Trials Objective Results: Number of Decide Requests Per Decision Per Minute Performance]{Number of Decide Requests Per Decision Per Minute.}
\begin{tabular}{c|c|c"c|c"c|c|}
\cline{2-7}
& \multicolumn{2}{c"}{Overall} & \multicolumn{2}{c"}{Easy} & \multicolumn{2}{c|}{Difficult} \\ \cline{2-7}
& \cellcolor{Gray}Mean & \cellcolor{Gray}Median & \cellcolor{Gray}Mean & \cellcolor{Gray}Median & \cellcolor{Gray}Mean & \cellcolor{Gray}Median \\
& (SD) & (Min/Max) & (SD) & (Min/Max) & (SD) & (Min/Max) \\ \thickhline
\multicolumn{1}{|c|}{\multirow{2}{*}{$M_{1}$}} & \cellcolor{Gray}0.1 & \cellcolor{Gray}0 & \cellcolor{Gray}0.12 & \cellcolor{Gray}0 & \cellcolor{Gray}0.08 & \cellcolor{Gray}0 \\
\multicolumn{1}{|c|}{} & (0.19) & (0/0.8) & (0.22) & (0/0.8) & (0.12) & (0/0.43) \\ \thickhline
\multicolumn{1}{|c|}{\multirow{2}{*}{$M_{2}$}} & \cellcolor{Gray}0.15 & \cellcolor{Gray}0.14 & \cellcolor{Gray}0.19 & \cellcolor{Gray}0.22 & \cellcolor{Gray}0.1 & \cellcolor{Gray}0 \\
\multicolumn{1}{|c|}{} & (0.16) & (0/0.55) & (0.18) & (0/0.55) & (0.13) & (0/0.41) \\ \thickhline
\multicolumn{1}{|c|}{\multirow{2}{*}{$M_{3}$}} & \cellcolor{Gray}0.23 & \cellcolor{Gray}0.21 & \cellcolor{Gray}0.26 & \cellcolor{Gray}0.24 & \cellcolor{Gray}0.2 & \cellcolor{Gray}0.17 \\
\multicolumn{1}{|c|}{} & (0.12) & (0/1.15) & (0.12) & (0.08/1.15) & (0.12) & (0/0.9) \\ \hline
\end{tabular}
\label{table:HC_Decide_Per_Decision_Time}
\end{table}

\begin{table}[bp!]
\centering
	\captionsetup{aboveskip=3pt}
	\caption[Human Trials Objective Results: Number of Abandon Requests Per Decision Per Minute Performance]{Number of Abandon Requests Per Decision Per Minute.}
\begin{tabular}{c|c|c"c|c"c|c|}
\cline{2-7}
& \multicolumn{2}{c"}{Overall} & \multicolumn{2}{c"}{Easy} & \multicolumn{2}{c|}{Difficult} \\ \cline{2-7}
& \cellcolor{Gray}Mean & \cellcolor{Gray}Median & \cellcolor{Gray}Mean & \cellcolor{Gray}Median & \cellcolor{Gray}Mean & \cellcolor{Gray}Median \\
& (SD) & (Min/Max) & (SD) & (Min/Max) & (SD) & (Min/Max) \\ \thickhline
\multicolumn{1}{|c|}{\multirow{2}{*}{$M_{1}$}} & \cellcolor{Gray}0.02 & \cellcolor{Gray}0 & \cellcolor{Gray}0.02 & \cellcolor{Gray}0 & \cellcolor{Gray}0.03 & \cellcolor{Gray}0 \\
\multicolumn{1}{|c|}{} & (0.09) & (0/0.6) & (0.08) & (0/0.6) & (0.1) & (0/0.5) \\ \thickhline
\multicolumn{1}{|c|}{\multirow{2}{*}{$M_{2}$}} & \cellcolor{Gray}0.02 & \cellcolor{Gray}0 & \cellcolor{Gray}0.02 & \cellcolor{Gray}0 & \cellcolor{Gray}0.02 & \cellcolor{Gray}0 \\
\multicolumn{1}{|c|}{} & (0.07) & (0/0.38) & (0.07) & (0/0.38) & (0.07) & (0/0.3) \\ \thickhline
\multicolumn{1}{|c|}{\multirow{2}{*}{$M_{3}$}} & \cellcolor{Gray}0.03 & \cellcolor{Gray}0 & \cellcolor{Gray}0.03 & \cellcolor{Gray}0 & \cellcolor{Gray}0.03 & \cellcolor{Gray}0 \\
\multicolumn{1}{|c|}{} & (0.07) & (0/0.36) & (0.07) & (0/0.36) & (0.07) & (0/0.35) \\ \hline
\end{tabular}
\label{table:HC_Abandon_Per_Dec_Time}
\end{table}

The type of request issued to the collectives characterizes the kind of influence the operators exerted.  Tables \ref{table:HC_Investigate_Per_Dec_Time}, \ref{table:HC_Decide_Per_Decision_Time}, and \ref{table:HC_Abandon_Per_Dec_Time} provide the number of investigate, abandon, and decide requests issued per decision per minute.  Investigate requests were the most common, given that the abandon and decide requests result in a persistent change to the collective's state and were required less frequently.  A Kruskal-Wallis test identified significant effects for investigate requests in overall, \textit{$\chi^{2}$(2, N = 1008) = 223.38, $\rho$ $<$ 0.001}, easy, \textit{$\chi^{2}$(2, N = 708) = 114.41, $\rho$ $<$ 0.001}, and difficult decisions, \textit{$\chi^{2}$(2, N = 550) = 111.64, $\rho$ $<$ 0.001}.  The success rates and investigate requests were weakly correlated for all decisions when using $M_{1}$, \textit{r = 0.36, $\rho$ $<$ 0.001}, and $M_{2}$, \textit{r = 0.11, $\rho$ = 0.04}, as well as difficult decisions using $M_{3}$, \textit{r = -0.19, $\rho$ = 0.03}. Investigate requests were strongly correlated with success rate for difficult decisions when using $M_{1}$, \textit{r = 0.62, $\rho$ $<$ 0.001}, indicating that greater human influence improved decision accuracy.  The number of investigate requests was significantly higher for $M_{3}$ than for the other models, but changes in Investigate Request Frequencies for both $M_{2}$ and $M_{3}$ did not correspond with changes in success rates with either model for easy, or difficult decisions. 

Decide requests occurred most frequently with $M_{3}$, as expected considering this model did not make independent decisions.  A Kruskal-Wallis test identified significant effects between models for decide requests in overall, \textit{$\chi^{2}$(2, N = 1008) = 164.34 $\rho$ $<$ 0.001}, easy, \textit{$\chi^{2}$(2, N = 569) = 80.24 $\rho$ $<$ 0.001}, and difficult decisions, \textit{$\chi^{2}$(2, N = 439) = 66.07 $\rho$ $<$ 0.001}.  A moderate positive correlation was observed for decide requests and success rate with $M_{1}$ in both overall, \textit{r = 0.2 $\rho$ $<$ 0.001}, and difficult, \textit{r = 0.36, $\rho$ $<$ 0.001}, decisions. A moderate, negative correlation was also observed for difficult decisions, \textit{r = -0.21, $\rho$ $<$ 0.01} with $M_{3}$.  The positive correlation for $M_{1}$ suggests that the operator often directed accurate decisions in difficult decisions.

Abandon requests were also less common than investigate requests, although a Kruskal-Wallis test revealed significant effects for abandon requests for overall, \textit{$\chi^{2}$(2, N = 1008) = 11.27, $\rho$ $<$ 0.01}, and easy decisions, \textit{$\chi^{2}$(2, N = 569) = 14.10, $\rho$ $<$ 0.001}.  A weak, negative correlation was observed between abandon request frequency and success rate for $M_{1}$ decisions overall, \textit{r = -0.13, $\rho$ = 0.02}, but no other correlations were identified.  

The differences between the models with respect to abandon request frequency was less informative than the differences observed in the intervention rate, which captures the abandon requests that were used to force a collective to ignore a target that had already gained at least 10\% of the population's support within the collective.  The descriptive statistics for the number of interventions per operator are presented in Table \ref{table:HC_Interventions_Per_Decision_Time}.  A Kruskal-Wallis test identified significant effects across the models for number of interventions in overall decisions, \textit{$\chi^{2}$(2, N = 84) = 10.35, $\rho$ $<$ 0.01}.  

\begin{table}[h!]
\centering
	\captionsetup{aboveskip=3pt}
	\caption[Human Trials Objective Results: Number of Interventions Per Participant]{Number of Interventions (Abandoned Targets with 10\% Support) Per Participant.}
\begin{tabular}{c|c|c|}
\cline{2-3}
& \multicolumn{2}{c|}{Overall} \\ \cline{2-3}
& \cellcolor{Gray}Mean & \cellcolor{Gray}Median \\
& (SD) & (Min/Max) \\ \thickhline
\multicolumn{1}{|c|}{\multirow{2}{*}{$M_{1}$}} & \cellcolor{Gray}4.79 & \cellcolor{Gray}4 \\
\multicolumn{1}{|c|}{} & (2.97) & (0/12) \\ \thickhline
\multicolumn{1}{|c|}{\multirow{2}{*}{$M_{2}$}} & \cellcolor{Gray}2.21 & \cellcolor{Gray}1.5 \\
\multicolumn{1}{|c|}{} & (1.99) & (0/7) \\ \thickhline
\multicolumn{1}{|c|}{\multirow{2}{*}{$M_{3}$}} & \cellcolor{Gray}5 & \cellcolor{Gray}3.5 \\
\multicolumn{1}{|c|}{} & (5.11) & (0/18) \\ \hline
\end{tabular}
\label{table:HC_Interventions_Per_Decision_Time}
\end{table}

These results indicate that although requests were less frequent with $M_{1}$ in general, the requests used were more likely to correspond to changes in the human-collective team's success rate than with the $M_{2}$ and $M_{3}$ models.  The positive correlations between $M_{1}$'s success rate and the number of requests, number of investigation requests, number of interventions, and number of decide requests suggests that in difficult decisions, the operator pushed the collectives towards accurate decisions.  Using available controls, the operators were able to take advantage of $M_{1}$'s speed during easy decisions, but were required to take control in difficult decisions in order to achieve a success rate similar to that achieved by $M_{2} SIM$, without human interaction, on difficult decisions.  Further, the high success rate observed with the baseline model, $M_{3}$, indicates that the operators were capable of driving the collectives to the correct decision, but these decisions took significantly longer to make and required many more operator actions to complete.

\begin{table}[bp!]
\centering
	\captionsetup{aboveskip=3pt}
	\caption[Human Trials Objective Results: Collective Observations Performance]{Collective Observations (\%) Per Decision.}
\begin{tabular}{c|c|c"c|c"c|c|}
\cline{2-7}
& \multicolumn{2}{c"}{Overall} & \multicolumn{2}{c"}{Easy} & \multicolumn{2}{c|}{Difficult} \\ \cline{2-7}
& \cellcolor{Gray}Mean & \cellcolor{Gray}Median & \cellcolor{Gray}Mean & \cellcolor{Gray}Median & \cellcolor{Gray}Mean & \cellcolor{Gray}Median \\
& (SD) & (Min/Max) & (SD) & (Min/Max) & (SD) & (Min/Max) \\ \thickhline
\multicolumn{1}{|c|}{\multirow{2}{*}{$M_{1}$}} & \cellcolor{Gray}77.98 & \cellcolor{Gray}100 & \cellcolor{Gray}65.62 & \cellcolor{Gray}100 & \cellcolor{Gray}94.44 & \cellcolor{Gray}100 \\
\multicolumn{1}{|c|}{} & (41.5) & (0/100) & (47.62) & (0/100) & (22.99) & (0/100) \\ \thickhline
\multicolumn{1}{|c|}{\multirow{2}{*}{$M_{2}$}} & \cellcolor{Gray}86.01 & \cellcolor{Gray}100 & \cellcolor{Gray}80 & \cellcolor{Gray}100 & \cellcolor{Gray}92.95 & \cellcolor{Gray}100 \\
\multicolumn{1}{|c|}{} & (34.74) & (0/100) & (40.11) & (0/100) & (25.68) & (0/100) \\ \thickhline
\multicolumn{1}{|c|}{\multirow{2}{*}{$M_{3}$}} & \cellcolor{Gray}90.18 & \cellcolor{Gray}100 & \cellcolor{Gray}88.32 & \cellcolor{Gray}100 & \cellcolor{Gray}92.81 & \cellcolor{Gray}100 \\
\multicolumn{1}{|c|}{} & (29.8) & (0/100) & (32.19) & (0/100) & (25.93) & (0/100) \\ \hline
\end{tabular}
\label{table:HC_Collective_Observations}
\end{table}

\begin{table}[bp!]
\centering
	\captionsetup{aboveskip=3pt}
	\caption[Human Trials Objective Results: Target Observations Performance]{Target Observations (\%) Per Decision.}
\begin{tabular}{c|c|c"c|c"c|c|}
\cline{2-7}
& \multicolumn{2}{c"}{Overall} & \multicolumn{2}{c"}{Easy} & \multicolumn{2}{c|}{Difficult} \\ \cline{2-7}
& \cellcolor{Gray}Mean & \cellcolor{Gray}Median & \cellcolor{Gray}Mean & \cellcolor{Gray}Median & \cellcolor{Gray}Mean & \cellcolor{Gray}Median \\
& (SD) & (Min/Max) & (SD) & (Min/Max) & (SD) & (Min/Max) \\ \thickhline
\multicolumn{1}{|c|}{\multirow{2}{*}{$M_{1}$}} & \cellcolor{Gray}20.54 & \cellcolor{Gray}0 & \cellcolor{Gray}10.94 & \cellcolor{Gray}0 & \cellcolor{Gray}33.33 & \cellcolor{Gray}0 \\
\multicolumn{1}{|c|}{} & (40.46) & (0/100) & (31.29) & (0/100) & (47.3) & (0/100) \\ \thickhline
\multicolumn{1}{|c|}{\multirow{2}{*}{$M_{2}$}} & \cellcolor{Gray}22.62 & \cellcolor{Gray}0 & \cellcolor{Gray}21.11 & \cellcolor{Gray}0 & \cellcolor{Gray}24.36 & \cellcolor{Gray}0 \\
\multicolumn{1}{|c|}{} & (41.9) & (0/100) & (40.92) & (0/100) & (43.06) & (0/100) \\ \thickhline
\multicolumn{1}{|c|}{\multirow{2}{*}{$M_{3}$}} & \cellcolor{Gray}41.07 & \cellcolor{Gray}0 & \cellcolor{Gray}41.12 & \cellcolor{Gray}0 & \cellcolor{Gray}41.01 & \cellcolor{Gray}0 \\
\multicolumn{1}{|c|}{} & (49.27) & (0/100) & (49.33) & (0/100) & (49.36) & (0/100) \\ \hline
\end{tabular}
\label{table:HC_Target_Observations}
\end{table}

\subsubsection{Situational Awareness}
\label{sec:SA_HC_Comparison}
 
Operators using $M_{1}$, $M_{2}$, and $M_{3}$, maintained varying levels of situational awareness while interacting with the collectives.  The descriptive statistics for operators' observation actions, including collective observations and target observations, are presented in Tables \ref{table:HC_Collective_Observations} and \ref{table:HC_Target_Observations}, respectively.  The displayed percentages are the ratio of the number of collective and target observation actions over the total actions for each decision.  The observation percentages differed significantly across the human-collective teams.  A Kruskal-Wallis test identified significant effects between the models for collective observations in overall, \textit{$\chi^{2}$(2, N = 1008) = 19.95, $\rho$ $<$ 0.001}, and easy decisions, \textit{$\chi^{2}$(2, N = 569) = 29.77, $\rho$ $<$ 0.001}.  Weak negative correlations were found between success rate and collective observations in overall decisions for $M_{1}$, \textit{r = -0.12, $\rho$ = 0.03}, and $M_{2}$, \textit{r = -0.12, $\rho$ = 0.03} (identical calculated values).  A Kruskal-Wallis test identified significant effects between the models for target observations per decision by overall, \textit{$\chi^{2}$(2, N = 1008) = 42.47, $\rho$ $<$ 0.001}, easy, \textit{$\chi^{2}$(2, N = 569) = 49.38, $\rho$ $<$ 0.001}, and difficult, \textit{$\chi^{2}$(2, N = 439) = 9.31, $\rho$ $<$ 0.01}, decisions.  Weak positive correlations were found between target observations and success rate with the $M_{3}$ teams for overall, \textit{r = 0.14, $\rho$ = 0.01}, and difficult, \textit{r = 0.16, $\rho$ = 0.05}, decisions.

The operators made the most collective observation actions with the baseline model, $M_{3}$, which was expected, although the number of collective observations during difficult decisions was similar for all the models.  Target observations were significantly more common with the baseline model, $M_{3}$ across decision difficulty.  Operators used fewer observation actions for easy decisions with $M_{1}$ when compared to $M_{2}$, but this relationship was reversed for difficult decisions.  The increased target observations for $M_{1}$ were not correlated with success rates for that model in difficult decisions, but they were more common, indicating that the operators were more likely to make several observations of $M_{1}$ in difficult decisions than they were for the same model in easier decisions.  Target observations were notably more common with $M_{3}$ than the other models.  As with the collective observations with $M_{1}$, target observations were also significantly more common with $M_{1}$ for difficult decisions than the corresponding easy decisions.  

The final objective metric observed were the operators' responses to the SA probe questions.  The SA probe responses evaluated the operators' understanding of the scenario, while interacting with the collectives.  The descriptive statistics for the SA Correct responses for each level of SA are provided in Table \ref{table:HC_SA_Responses} \cite{Roundtree20191}.  No significant effects were observed between the models for overall SA probe response accuracy, which were above 85\% overall.  The response accuracy was lowest for $SA_{3}$, which required the operators to forecast a model's future behavior (see Section \ref{sec:Experimental_Design_Best_of_N}). A Kruskal-Wallis test identified significant effects between the models for $SA_{3}$ probe questions, \textit{$\chi^{2}$(2, N = 84) = 7.57, $\rho$ = 0.02}.  The operators achieved a significantly higher SA Level 3 correct response rate when using $M_{2}$, as compared to the other models.  Across the SA levels within each model, the correct response rate differences were significant. Kruskal-Wallis tests also identified significant effects between levels for each model: $M_{1}$ - \textit{$\chi^{2}$(2, N = 56) = 64.63, $\rho$ $<$ 0.001}, $M_{2}$ - \textit{$\chi^{2}$(2, N = 56) = 73.68, $\rho$ $<$ 0.001}, and $M_{3}$ - \textit{$\chi^{2}$(2, N = 56) = 52.337, $\rho$ $<$ 0.001}.  The human-collective team correct response percentage dropped more than 8\% between $SA_{1}$ and $SA_{3}$ when using $M_{1}$ and 18\% for $M_{3}$.  The human-collective teams using $M_{2}$, in contrast, experienced less than a 2\% reduction between $SA_{1}$ and $SA_{3}$.  The SA probe correct response percentages demonstrate that the human operators maintained a consistent ability across all models to access information about the collectives ($SA_{1}$), and understand the collectives' processes ($SA_{2}$).  The $SA_{3}$ correct response percentages suggest that the use of the $M_{2}$ model either improved the human operator's ability to forecast the collective state, or afforded the human operator more opportunities to properly respond.

\begin{table}[h!]
\centering
	\captionsetup{aboveskip=3pt}
	\caption[Human Trials Subjective Results: SA Probe]{Percent Correct SA Probe Questions by SA Level.}
\begin{tabular}{c|c|c"c|c"c|c"c|c|}
\cline{2-9}
& \multicolumn{2}{c"}{$SA_{0}$} & \multicolumn{2}{c"}{$SA_{1}$} & \multicolumn{2}{c"}{$SA_{2}$} & \multicolumn{2}{c|}{$SA_{3}$}\\ \cline{2-9}
& \cellcolor{Gray}Mean & \cellcolor{Gray}Median & \cellcolor{Gray}Mean & \cellcolor{Gray}Median & \cellcolor{Gray}Mean & \cellcolor{Gray}Median & \cellcolor{Gray}Mean & \cellcolor{Gray}Median \\
& (SD) & (Min/Max) & (SD) & (Min/Max) & (SD) & (Min/Max) & (SD) & (Min/Max) \\ \thickhline
\multicolumn{1}{|c|}{\multirow{2}{*}{$M_{1}$}} & \cellcolor{Gray}85.12 & \cellcolor{Gray}91.67 & \cellcolor{Gray}86.31 & \cellcolor{Gray}100 & \cellcolor{Gray}87.32 & \cellcolor{Gray}100 & \cellcolor{Gray}78.57 & \cellcolor{Gray}100 \\
\multicolumn{1}{|c|}{} & (17.18) & (8.33/100) & (21.06) & (0/100) & (18.23) & (25/100) & (31.38) & (0/100) \\ \thickhline
\multicolumn{1}{|c|}{\multirow{2}{*}{$M_{2}$}} & \cellcolor{Gray}89.88 & \cellcolor{Gray}91.67 & \cellcolor{Gray}91.67 & \cellcolor{Gray}100 & \cellcolor{Gray}88.39 & \cellcolor{Gray}100 & \cellcolor{Gray}89.88 & \cellcolor{Gray}100 \\
\multicolumn{1}{|c|}{} & (10.96) & (58.33/100) & (11.11) & (66.67/100) & (14.6) & (60/100) & (20.46) & (33.33/100) \\ \thickhline
\multicolumn{1}{|c|}{\multirow{2}{*}{$M_{3}$}} & \cellcolor{Gray}87.2 & \cellcolor{Gray}91.67 & \cellcolor{Gray}94.05 & \cellcolor{Gray}100 & \cellcolor{Gray}91.43 & \cellcolor{Gray}100 & \cellcolor{Gray}76.79 & \cellcolor{Gray}75 \\
\multicolumn{1}{|c|}{} & (10.75) & (58.33/100) & (13) & (66.67/100) & (12.68) & (60/100) & (16.57) & (50/100) \\ \hline
\end{tabular}
\label{table:HC_SA_Responses}
\end{table}

The situational awareness results demonstrate that the operators needed to take many more observation actions with $M_{3}$ than the collective action selection models.  Across decision difficulty the observation activity of the $M_{2}$ and $M_{3}$ teams were consistent, but the observation behavior for the $M_{1}$ team noticeably increased between easy and difficult decisions.  The SA probe correct response rates were higher than anticipated for all three models, which indicates that the Collective Interface Visualization was sufficient to enable operator situational awareness.  The consistently better $SA_{3}$ response percentages for human-collective teams using $M_{2}$ strongly suggests an advantage of this model over the other two for the human-operator.  Due to the simultaneous decision-making during the trials, associating individual SA probe questions to decision problem difficulty was not possible; however, these results suggest that future examination of SA probe responses under different decision difficulties is likely to further distinguish these models.

\begin{table}[bp!]
\centering
	\captionsetup{aboveskip=3pt}
	\caption[Human Trials Subjective Results: SART]{SART Results for Overall Score, Situational Understanding (SU), Demand on Attentional Resources (DAR), and Supply of Attentional Resources (SAR) with Overall = SU - (DAR - SAR).}
\begin{tabular}{c|c|c"c|c"c|c"c|c|}
\cline{2-9}
& \multicolumn{2}{c"}{Overall Score} & \multicolumn{2}{c"}{SU} & \multicolumn{2}{c"}{DAR} & \multicolumn{2}{c|}{SAR}\\ \cline{2-9}
& \cellcolor{Gray}Mean & \cellcolor{Gray}Median & \cellcolor{Gray}Mean & \cellcolor{Gray}Median & \cellcolor{Gray}Mean & \cellcolor{Gray}Median & \cellcolor{Gray}Mean & \cellcolor{Gray}Median \\
& (SD) & (Min/Max) & (SD) & (Min/Max) & (SD) & (Min/Max) & (SD) & (Min/Max) \\ \thickhline
\multicolumn{1}{|c|}{\multirow{2}{*}{$M_{1}$}} & \cellcolor{Gray}6 & \cellcolor{Gray}6 & \cellcolor{Gray}5.75 & \cellcolor{Gray}6 & \cellcolor{Gray}5 & \cellcolor{Gray}5 & \cellcolor{Gray}5.25 & \cellcolor{Gray}5.5 \\
\multicolumn{1}{|c|}{} & (2.28) & (1/13) & (1) & (3/7) & (1.22) & (1/7) & (1.48) & (2/7) \\ \thickhline
\multicolumn{1}{|c|}{\multirow{2}{*}{$M_{2}$}} & \cellcolor{Gray}6.68 & \cellcolor{Gray}6.5 & \cellcolor{Gray}6.07 & \cellcolor{Gray}6 & \cellcolor{Gray}5.07 & \cellcolor{Gray}5 & \cellcolor{Gray}5.68 & \cellcolor{Gray}6 \\
\multicolumn{1}{|c|}{} & (2.26) & (3/13) & (0.9) & (4/7) & (1.18) & (1/6) & (1.09) & (3/7) \\ \thickhline
\multicolumn{1}{|c|}{\multirow{2}{*}{$M_{3}$}} & \cellcolor{Gray}6.39 & \cellcolor{Gray}6 & \cellcolor{Gray}6.07 & \cellcolor{Gray}6 & \cellcolor{Gray}5.04 & \cellcolor{Gray}5 & \cellcolor{Gray}5.36 & \cellcolor{Gray}5 \\
\multicolumn{1}{|c|}{} & (2.08) & (4/11) & (0.98) & (4/7) & (1.43) & (1/7) & (1.31) & (3/7) \\ \hline
\end{tabular}
\label{table:HC_SART}
\end{table}

\subsubsection{Subjective Results}
\label{sec:Subjective_HC_Comparison}

\begin{table}[bp!]
\centering
	\captionsetup{aboveskip=3pt}
	\caption[Human Trials Subjective Results: NASA-TLX]{Significant NASA-TLX results for Overall Score, Mental Demand and Temporal Demand.  No signficant effects observed for Physical, Performance, Effort and Frustration NASA --TLX Components (omitted).}
\begin{tabular}{c|c|c"c|c"c|c|}
\cline{2-7}
& \multicolumn{2}{c"}{Overall} & \multicolumn{2}{c"}{Mental} & \multicolumn{2}{c|}{Temporal} \\ \cline{2-7}
& \cellcolor{Gray}Mean & \cellcolor{Gray}Median &\cellcolor{Gray}Mean & \cellcolor{Gray}Median & \cellcolor{Gray}Mean & \cellcolor{Gray}Median  \\
& (SD) & (Min/Max) & (SD) & (Min/Max) & (SD) & (Min/Max) \\ \thickhline
\multicolumn{1}{|c|}{\multirow{2}{*}{$M_{1}$}} & \cellcolor{Gray}58.31 & \cellcolor{Gray}60.67 &  \cellcolor{Gray}22.19 & \cellcolor{Gray}23.33 & \cellcolor{Gray}11.55 & \cellcolor{Gray}9.67  \\
\multicolumn{1}{|c|}{} & (17.63) & (9/89.33) & (6.38) & (1.00/31.67) & (8.41) & (0.00/28.33) \\ \thickhline
\multicolumn{1}{|c|}{\multirow{2}{*}{$M_{2}$}} & \cellcolor{Gray}57.06 & \cellcolor{Gray}56.83 & \cellcolor{Gray}23.58 & \cellcolor{Gray}25.00 & \cellcolor{Gray}10.94 & \cellcolor{Gray}10.33\\
\multicolumn{1}{|c|}{} & (16.47) & (5.67/83.33) & (6.28) & (3.00/31.67)
& (7.60) & (0.00/24.00)\\ \thickhline
\multicolumn{1}{|c|}{\multirow{2}{*}{$M_{3}$}} & \cellcolor{Gray}50.63 & \cellcolor{Gray}54.17 & \cellcolor{Gray}16.54 & \cellcolor{Gray}18.50 & \cellcolor{Gray}7.49 & \cellcolor{Gray}5.33\\
\multicolumn{1}{|c|}{} & (17.56) & (9.33/80.33) & (9.10) & (0.00/33.33) & (6.57) & (0.00/22.67) \\ \hline
\end{tabular}
\label{table:HC_NASA_TLX}
\end{table}

This section presents the subjective data including the operator's reported situational awareness, workload,  post-trial questionnaires, post-experiment questionnaires, and the Mental Rotation Tests.  The descriptive statistics for the 3-D SART and NASA-TLX are presented in Tables \ref{table:HC_SART} and \ref{table:HC_NASA_TLX} \cite{Roundtree20191}, respectively.  A Kruskal-Wallis test did not reveal significant effects between the models for the overall 3-D SART score, or any of the 3-D SART components.  The scores were similar across the models, but slightly higher for $M_{2}$.   

A Kruskal-Wallis revealed no significant effects for the overall NASA-TLX score, but identified significant effects between the models for mental demand, \textit{$\chi^{2}$(2, N = 84) = 22.166, $\rho$ $<$ 0.001}, and temporal demand, \textit{$\chi^{2}$(2, N = 84) = 8.8327, $\rho$ = 0.012}.  A pairwise  comparison using a Tukey and Kramer test revealed significant effects between $M_{1}$ and $M_{3}$  (\textit{$\rho$ = 0.003}) and $M_{2}$ and $M_{3}$ (\textit{$\rho$ $<$ 0.001}) for mental demand.  A similar test revealed significant effects for Temporal Demand between $M_{1}$ and $M_{3}$ (\textit{$\rho$ = 0.023}) and between $M_{2}$ and $M_{3}$ (\textit{$\rho$ = 0.032}).  The operators reported higher mental and temporal demand when using models $M_{1}$ and $M_{2}$.  Higher mental demand indicates higher required perceptual or decision making activity.  The higher mental demand reported for $M_{1}$ and $M_{2}$ suggests that operators experienced higher demand when sharing decision making tasks with the other collectives.  When making decisions, as in $M_{3}$, the operators did not need to consider what the collective was doing.  The higher temporal demand is likely due to the fact that both models, $M_{1}$ and $M_{2}$, made independent decisions, whether the human operator intervened or not.  The use of these models likely introduced additional pressure on the human operator to move quickly in order to influence each collective's decisions, before the collective made an independent decision.  $M_{3}$ did not impose a similar pressure for the human operator to act.    

The post-trial questionnaires required Likert scale responses, on a scale of 1 (worst) to 7 (best) regarding the effectiveness of each type of request, the collective model's responsiveness to requests, the collective model's independent target selection ability (Performance), and ease of understanding (Comprehension).   The descriptive statistics are shown in Tables \ref{table:HC_post_trial_response} and \ref{table:HC_post_trial_performance_understanding}, respectively.  A Kruskal-Wallis test identified significant effects between the models for the effectiveness of the Abandon request, \textit{$\chi^{2}$(2, N = 84) = 6.33, $\rho$ = 0.04}, and the model's independent performance, \textit{$\chi^{2}$(2, N = 84) = 6.8, $\rho$ = 0.03}.  The fact that the operators rated the effectiveness of the Abandon request lowest for the baseline model is not surprising, since the model only persistently investigated targets dictated by the human.  The low performance reported for $M_{1}$ is consistent with the lower performance of the independent model, $M_{1} SIM$, compared to $M_{2} SIM$ and the lower performance of the $M_{1}$ human-collective teams compared to the $M_{2}$ human collective teams. 

\begin{table}[bp!]
\centering
	\captionsetup{aboveskip=3pt}
	\caption[Human Trials Subjective Results: Post Trial Request Evaluation]{Post Trial Request Type Effectiveness Ranking (1-low, 7-high).}
\begin{tabular}{c|c|c"c|c"c|c|}
\cline{2-7}
& \multicolumn{2}{c"}{Investigate} & \multicolumn{2}{c"}{Abandon} & \multicolumn{2}{c|}{Decide} \\ \cline{2-7}
& \cellcolor{Gray}Mean & \cellcolor{Gray}Median & \cellcolor{Gray}Mean & \cellcolor{Gray}Median & \cellcolor{Gray}Mean & \cellcolor{Gray}Median \\
& (SD) & (Min/Max) & (SD) & (Min/Max) & (SD) & (Min/Max) \\ \thickhline
\multicolumn{1}{|c|}{\multirow{2}{*}{$M_{1}$}} & \cellcolor{Gray}5.07 & \cellcolor{Gray}5 & \cellcolor{Gray}6.04 & \cellcolor{Gray}6 & \cellcolor{Gray}5.71 & \cellcolor{Gray}7 \\
\multicolumn{1}{|c|}{} & (1.25) & (2/7) & (1.35) & (1/7) & (1.94) & (1/7) \\ \thickhline
\multicolumn{1}{|c|}{\multirow{2}{*}{$M_{2}$}} & \cellcolor{Gray}4.75 & \cellcolor{Gray}5 & \cellcolor{Gray}6.18 & \cellcolor{Gray}7 & \cellcolor{Gray}5.57 & \cellcolor{Gray}6 \\
\multicolumn{1}{|c|}{} & (1.53) & (2/7) & (1.42) & (1/7) & (1.99) & (1/7) \\ \thickhline
\multicolumn{1}{|c|}{\multirow{2}{*}{$M_{3}$}} & \cellcolor{Gray}5.18 & \cellcolor{Gray}5.5 & \cellcolor{Gray}5.29 & \cellcolor{Gray}5.5 & \cellcolor{Gray}6.54 & \cellcolor{Gray}7 \\
\multicolumn{1}{|c|}{} & (1.68) & (1/7) & (1.76) & (1/7) & (0.92) & (4/7) \\ \hline
\end{tabular}
\label{table:HC_post_trial_response}
\end{table}

\begin{table}[bp!]
\centering
	\captionsetup{aboveskip=3pt}
	\caption[Human Trials Subjective Results: Post Trial Performance and Understanding]{Post Trial Performance and Understanding Model Ranking (1-low, 7-high).}
\begin{tabular}{c|c|c"c|c|c|c|}
\cline{2-5}
& \multicolumn{2}{c"}{Performance} & \multicolumn{2}{c|}{Understanding} \\ \cline{2-5}
& \cellcolor{Gray}Mean & \cellcolor{Gray}Median & \cellcolor{Gray}Mean & \cellcolor{Gray}Median \\
& (SD) & (Min/Max) & (SD) & (Min/Max) \\ \thickhline
\multicolumn{1}{|c|}{\multirow{2}{*}{$M_{1}$}} & \cellcolor{Gray}5.11 & \cellcolor{Gray}5 & \cellcolor{Gray}5.39 & \cellcolor{Gray}5.5 \\
\multicolumn{1}{|c|}{} & (0.99) & (3/7) & (1.31) & (2/7) \\ \thickhline
\multicolumn{1}{|c|}{\multirow{2}{*}{$M_{2}$}} & \cellcolor{Gray}5.54 & \cellcolor{Gray}6 & \cellcolor{Gray}5.82 & \cellcolor{Gray}6 \\
\multicolumn{1}{|c|}{} & (1.29) & (3/7) & (1.16) & (3/7) \\ \thickhline
\multicolumn{1}{|c|}{\multirow{2}{*}{$M_{3}$}} & \cellcolor{Gray}5.75 & \cellcolor{Gray}6 & \cellcolor{Gray}5.93 & \cellcolor{Gray}7 \\
\multicolumn{1}{|c|}{} & (1.43) & (2/7) & (1.46) & (3/7) \\ \hline
\end{tabular}
\label{table:HC_post_trial_performance_understanding}
\end{table}

\begin{table}[bp!]
\centering
	\captionsetup{aboveskip=3pt}
	\caption[Human Trials Subjective Results: Post Experiment Evaluation]{Post Experiment Model Ranking (1-best, 3-worst).}
\begin{tabular}{c|c|c"c|c"c|c|}
\cline{2-7}
& \multicolumn{2}{c"}{Responsiveness} & \multicolumn{2}{c"}{Performance} & \multicolumn{2}{c|}{Comprehension} \\ \cline{2-7}
& \cellcolor{Gray}Mean & \cellcolor{Gray}Median & \cellcolor{Gray}Mean & \cellcolor{Gray}Median & \cellcolor{Gray}Mean & \cellcolor{Gray}Median \\
& (SD) & (Min/Max) & (SD) & (Min/Max) & (SD) & (Min/Max) \\ \thickhline
\multicolumn{1}{|c|}{\multirow{2}{*}{$M_{1}$}} & \cellcolor{Gray}1.5 & \cellcolor{Gray}1.5 & \cellcolor{Gray}2 & \cellcolor{Gray}2 & \cellcolor{Gray}2.5 & \cellcolor{Gray}2.5 \\
\multicolumn{1}{|c|}{} & (0.51) & (1/2) & (1.02) & (1/3) & (0.51) & (2/3) \\ \thickhline
\multicolumn{1}{|c|}{\multirow{2}{*}{$M_{2}$}} & \cellcolor{Gray}1.5 & \cellcolor{Gray}1.5 & \cellcolor{Gray}2 & \cellcolor{Gray}2 & \cellcolor{Gray}2.5 & \cellcolor{Gray}2.5 \\
\multicolumn{1}{|c|}{} & (0.51) & (1/2) & (1.02) & (1/3) & (0.51) & (2/3) \\ \thickhline
\multicolumn{1}{|c|}{\multirow{2}{*}{$M_{3}$}} & \cellcolor{Gray}3 & \cellcolor{Gray}3 & \cellcolor{Gray}2 & \cellcolor{Gray}2 & \cellcolor{Gray}1 & \cellcolor{Gray}1 \\
\multicolumn{1}{|c|}{} & (0) & (3/3) & (0) & (2/2) & (0) & (1/1) \\ \hline
\end{tabular}
\label{table:HC_post_experiment_ranking}
\end{table}

The post experiment questionnaire required the operators to rank order the different models according to overall Responsiveness to requests, overall performance, and overall Comprehension.  The descriptive statistics for the post experimental rankings are summarized in Table \ref{table:HC_post_experiment_ranking}.  A Kruskal-Wallis test revealed significant effects for Responsiveness, \textit{$\chi^{2}$(2, N = 28) = 62.25, $\rho$ $<$ 0.001}, and Comprehension, \textit{$\chi^{2}$(2, N = 28) = 62.25, $\rho$ $<$ 0.001} (identical values).  The ranking of the baseline model, $M_{3}$ overall is interesting, as it was ranked consistently the lowest for responsiveness, the highest for comprehension, and between the remaining models for performance.  

The final subjective metric was the Mental Rotations Test scores.  The MRT has a minimum score of $0$ and a maximum possible score of 24.  The operators' MRT scores resulted in a 10.9 mean score (standard deviation = $\pm 5.5$, median = 10, minimum = 1, maximum = 24) \cite{Roundtree20191}.  These results were virtually identical to the results of a large study comprised of 636 operators with a 10.8 mean score and a standard deviation of $\pm 5$ \cite{MRT_peters1995redrawn}.  The results of the MRT were compared to the human-collective team success rates for each problem type.  A Spearman correlations test only identified weak to moderate positive correlations between the MRT results and success rates with model $M_{1}$ for overall, \textit{r = 0.2, $\rho$ $<$ 0.001, easy, r = 0.15, $\rho$ = 0.03}, and difficult decisions, \textit{r = 0.36, $\rho$ $<$ 0.001}.  These findings suggest that operators with higher spatial awareness were slightly more likely to make accurate decisions with the $M_{1}$ collective action selection model.

\section{Discussion}
\label{sec:HumanCollectiveDiscussion}

The results of the experiments supported the overall research goal.  The bias reducing mechanisms enabled the collective to make better decisions with, or without, a human operator when compared to the model without these mechanisms.  The Collective Interface Visualization and interaction strategy were sufficient for the human operator to maintain awareness of the collectives' states and influence the best-of-$n$ decision process in order to improve human-collective team's decision accuracy and time.  When using the bias reducing mechanisms, the interaction strategy demonstrated significant fault tolerance \cite{HSI_SharedControl}, which means the team's success was resilient to changes in human operator behavior.  The reliance of the human-collective team on the actions of the human operator differed depending upon whether or not the collective action selection model used bias reducing mechanisms.  When using model $M_{1}$, the success of the human-collective team required the human operator to take control during difficult decisions in order to overcome the model's susceptibility to negative environmental bias.  When using $M_{2}$, the human-collective team made accurate decisions with relative indifference to the frequency of human operator actions.       

The independent collective experiment supported hypothesis $H_{1}$, which predicted that the bias reduction mechanisms of $M_{2}SIM$ improves decision accuracy when compared to the original model, $M_{1}SIM$.  The results provide support of the well known trade-off between speed and accuracy in these types of decisions \cite{Valentini2015:EfficientDecisionMakingSpeedvsAccuracy}. The original model is fast when making decisions in which the environmental bias favors the best site (e.g., the best site is closest or equidistant to other sites).  When required to select the best target,  regardless of the target's distance from a collective's decision making hub, $M_{1} SIM$ is unreliable and makes poor decisions in difficult decision problems.  $M_{2} SIM$'s resistance to environmental bias is a significant advantage over the original model, especially when accuracy in the best-of-$n$ decision is more important than the time required to make a decision.  Future investigations will determine if other bias reduction mechanisms reduce the decision times achieved by $M_{2} SIM$  (e.g., modulating interaction rates over time rather than distance \cite{ReinaICRA:2019}).  

Hypothesis, $H_{2}$, which predicted that human influence improves the decision processes of the collective action selection models, was supported.  The human-collective teams ($M_{1}$ and $M_{2}$) made faster, more accurate decisions than the independent collectives ($M_{1}SIM$ and $M_{2}SIM$).  The human-collective teams using the baseline model, $M_{3}$, were very accurate, but required the most human operator actions and made the slowest decisions of all the human-collective teams.  Comparing the behavior of the $M_{1}$ and $M_{2}$ human-collective teams reveals an important distinction between the two collective action selection models.  Unlike the human-collective teams using $M_{2}$, the $M_{1}$ teams' success rates were more strongly correlated with the human operators' request frequencies, use of investigate requests, use of decide requests, and spatial awareness (determined using MRT results).  Further, the human operators were less than half as likely to intervene in the collective's decisions when using $M_{2}$ than when they were using $M_{1}$, which may have occurred in order to correct $M_{1}$'s faster, but less accurate decision making.   The lack of similar observations with the $M_{2}$ human-collective teams supports that the $M_{2}$ model is significantly more tolerant to changes in human operator behavior than the $M_{1}$ model.  Despite the objective results, the operators struggled to distinguish the models subjectively, despite correcting the collectives' decision making process more frequently with $M_{1}$ and achieving higher decision accuracy with $M_{2}$.

The final hypothesis, $H_{3}$, which stated that the Collective Interface Visualization and interaction strategy were sufficient to enable human interaction with the collective action selection models, was also supported.  The operators' influence increased the selection of the highest valued targets, when compared to the collective decision processes alone ($M_{1}SIM$ and $M_{2}SIM$).  The operators' use of investigate and abandon requests to slow down the $M_{1}$ decisions and speed up the $M_{2}$ decisions demonstrates operator control over both collective action selection models in order to improve the human-collective team's decision accuracy and decision time.  The operators responded consistently well to the $SA_{1}$ (perception) and $SA_{2}$ (comprehension) probe questions, suggesting that the operators perceived and understood the collectives' states.  The ability to forecast a collective's future state ($SA_{3}$) was significantly lower for the human-collective teams when using models $M_{1}$ and $M_{3}$.  Operators may have been overloaded with the rapid change of collective behaviors and necessary adjustments to improve decision accuracy when using $M_{1}$, while $M_{3}$ may have preoccupied operators with the amount of influence required in the decision making process.  When using $M_{2}$, the human operators were either afforded greater opportunity, or capability to answer the $SA_{3}$ questions correctly.  The objective observations contrast with the reported NASA-TLX workload, in which operators reported relatively equal mental and temporal demand between $M_{1}$ and $M_{2}$, but the lowest demands with $M_{3}$.  This contrast between the objective and subjective measures is partially explained for the baseline model by the experimental design.  Operator experience during the two trials preceding the $M_{3}$ trial likely positively influenced some of the subjective measures, including workload and model comparisons, due to learning affects.  Additionally, since human operators controlled collective decision making states when using model $M_{3}$, they were expected to better understand the model's state.  A challenge imposed by the experimental design was the inability to categorize subjective measures between difficult and easy decisions.  Most of the objective measures differed between the decision difficulty, but the number of interventions, as well as the NASA-TLX, 3-D SART, and operator evaluations were not assessed by decision difficulty.  

Quantifying the Collective Interface Visualization's effectiveness requires its comparison with other visualization strategies. The Collective Interface Visualization enabled the operators to interact with each collective as a single entity, rather than as organizations of two hundred independent entities.  Operators often observed a collective's decision-making state by reviewing the percentage of collective individuals performing each available behavior and the support visible in each collective's considered targets. An operator's requests to a collective's individuals were limited to the collective hub, with a report confirming to the operator that a request had been issued properly. The results demonstrate that operators consistently, and effectively, observed the collective state and influenced the collectives' decision-making processes.

Some design limitations of the Collective Interface Visualization can be improved in order to promote better human-collective behaviors.  The inability to observe individual collective entity activities and the immediate changes in collective behavior after executing particular requests, may have caused operators to reissue requests.  The messages and status indicators (i.e., circular colored icons) provided in the Collective Assignments area were intended to communicate information about operator requests. The information may have not been salient enough to attract the operator's attention to the Collective Assignments area to confirm that their requests were executed or still in progress.  Dividing the Collective Assignments area into two sections, one explicitly for executed requests and another for ongoing requests, instead of interpreting the colored icons, may help improve operator comprehension of a request's status. Alternatively, the words ``executed" highlighted in red, or ``in progress" highlighted in green, can be added to each request line drawing attention through the use of color and providing redundancy through text.  The effectiveness of these design changes are only useful assuming the operators are not colorblind, as was done in this evaluation.  The difference between the collective action selection models may be more perceivable if directional information, in the form of arrows or streamlines emitting from the Collective's hub, indicated which targets were being favored.  Operators may be able to see the bias towards nearby targets visually using the $M_{1}$ model versus the consistent favoring of high valued targets, irrespective of the distance to the hub, using the $M_{2}$ model.  Future evaluations are needed in order to affirm whether design changes to the Collective Interface Visualization improve human-collective behaviors, or if alternative visualization strategies are more effective.

Despite these limitations, the results are promising.  The presented bias reducing mechanisms embedded into the $M_{2}$ model enabled more accurate collective decisions.  Human-collective teams using the bias reducing model were more accurate and relied less on human influenc, than the teams that did not use these models.  Finally, an abstract visualization that aggregates collective state information enabled the human operators to successfully supervise and influence the decision making of multiple collectives simultaneously.

\section{Conclusion}
\label{sec:Conclusion}
This manuscript evaluated a human-collective strategy intended to improve the collective's ability to inform human operator decisions and the human operator's ability to supervise and influence the collective's behavior.  Two novel collective action selection models were evaluated in order to determine the effectiveness of new bias reduction mechanisms on the collectives' decisions, the ability of human operators to improve the collectives' decision making, and the feasibility of a novel human-collective interface. The bias-reducing model made more accurate, but slower, collective best-of-$n$ decisions than the model without these mechanisms in the absence of human-operator influence.  Human influence improved the collective decision accuracy and reduced decision times over the collectives alone for both action selection models.  The collective without the bias-reduction required human operator influence in order to drive the human-collective team to make accurate decisions.  Further, the evaluated Collective Interface was sufficient to enable a human operator to influence four collectives, comprised of 200 individuals each, simultaneously solving the sequential best-of-$n$ problem in static targeting scenarios.  Although promising, this manuscript highlights the need for future work in order to examine different visualization strategies, alternative bias reducing mechanisms, or other human-collective control mechanisms in order to realize the benefits of human-collective shared decision making in real world environments.

\section{Acknowledgments}
\label{sec:Acknowledgments}

 This work was partially funded by the US Office of Naval Research Awards N000141210987, N00014161302, N000141613025.  The work of Jason R. Cody was supported by the United States Military Academy and the United States Army Advanced Civil Schooling (ACS) program.  The views and conclusions contained herein are those of the authors and is not to be interpreted as necessarily representing the official policies or endorsements, either expressed or implied, of the United States Military Academy, the U.S. Army, or the U.S. Government.

\bibliographystyle{ACM-Reference-Format}
\bibliography{Library}

\end{document}